\documentclass[]{spie}  %>>> use for US letter paper
\usepackage{amsmath,amsfonts,amssymb}
\usepackage{graphicx}
\usepackage[colorlinks=true, allcolors=blue]{hyperref}

\usepackage{float}
\usepackage{caption}
\usepackage{subcaption}
\usepackage{textcomp}
\usepackage{array}

\title{Optical design of VIPER: a high-resolution multimode fiber-fed VIPA spectrograph for characterizing\\exoplanet atmospheric escape}

\author[a,b]{Matthew C. H. Leung}
\author[a,b]{David Charbonneau}
\author[a,b,c]{Andrew Szentgyorgyi}
\author[a,c]{\newline Colby Jurgenson}

\affil[a]{Center for Astrophysics \textbar{} Harvard \& Smithsonian, 60 Garden St, Cambridge, MA 02138, USA}
\affil[b]{Department of Astronomy, Harvard University, 60 Garden St, Cambridge, MA 02138, USA}
\affil[c]{Smithsonian Astrophysical Observatory, 100 Acorn Park Dr, Cambridge, MA 02140, USA}

\authorinfo{Further author information: (Send correspondence to M.C.H. Leung)\\M.C.H. Leung: E-mail: matthew.leung@cfa.harvard.edu}

\begin{document} 
\maketitle

\begin{abstract}
We present the optical design of VIPER: a high-resolution, multimode fiber-fed, narrowband, cross-dispersed, seeing-limited spectrograph based on a Virtually Imaged Phased Array (VIPA), a spectral disperser that provides higher dispersion and a more compact form factor than conventional diffraction gratings. VIPER is specifically designed to probe exoplanet atmospheric escape through the helium 1083 nm triplet line, with a resolving power of 300,000 over a 10 nm wavelength range. VIPER is intended for operation at the 1.5~m Tillinghast Telescope at the Fred Lawrence Whipple Observatory (FLWO) on Mount Hopkins, Arizona, USA, and is matched to this telescope's 100~\textmu m circular-core multimode optical fiber feed at f/6. VIPER's optical design is specifically optimized for high throughput, which, in a VIPA spectrograph, is more challenging with a multimode fiber feed than a single-mode fiber feed because of the larger etendue. To address this challenge, our design implements a novel use of a cylindrical beam expander and pupil slicer. We validate our design using analytic calculations and simulations in Zemax OpticStudio non-sequential mode. Our results demonstrate the feasibility of multimode fiber-fed VIPA spectrographs as compact, high-throughput alternatives to conventional grating-based spectrographs for exoplanet science and other astronomical applications.
\end{abstract}


%%%%%%%%%%%%%%%%%%%%%%%%%%%%%%%%%%%%%%%%%%%%%%%%%%%%%%%%%%%%%%%%%%%%
%%%%%%%%%%%%%%%%%%%%%%%%%%%%%%%%%%%%%%%%%%%%%%%%%%%%%%%%%%%%%%%%%%%%

\section{Introduction}

High-resolution spectrographs are crucial in many areas of astronomy. However, conventional grating-based astronomical spectrographs scale poorly with increased resolving power, typically becoming larger\cite{Schroeder} and more expensive. One compelling alternative that has recently gained interest is astronomical spectrographs based on a Virtually Imaged Phased Array (VIPA\cite{Shirasaki1996}), a relatively new spectral disperser that provides higher dispersion and throughput compared to a diffraction grating, in a more compact form factor. VIPA spectrographs are particularly well-suited for applications requiring high resolution and high throughput, without the need for a broad spectral range. An increasing number of applications in exoplanet science align well with this.

We are developing VIPER\cite{Leung2025}, a high-resolution, multimode fiber-fed, narrowband, seeing-limited VIPA spectrograph for the 1.5 m Tillinghast Telescope at the Fred Lawrence Whipple Observatory (FLWO) on Mount Hopkins, Arizona, USA. VIPER is specifically designed to probe exoplanet atmospheric escape through the metastable helium 1083 nm triplet line, with a resolving power of 300,000 over a 10 nm wavelength range. In this work, we present the most recent iteration of VIPER's optical design, which is optimized for high throughput. In particular, we discuss the challenges associated with using a multimode optical fiber, instead of a single-mode optical fiber, to feed a VIPA spectrograph, and how we mitigated these challenges.

An outline of this paper is as follows. In Section~\ref{sec:VIPA_spec_overview}, we provide an overview of how a VIPA and a cross-dispersed VIPA spectrograph function, and we discuss the challenges of using a multimode fiber to feed a VIPA spectrograph. In Section~\ref{sec:inst_overview}, we provide a description of VIPER's science drivers, instrument requirements, and an instrument overview. In Section~\ref{sec:optical_design}, we present VIPER's optical design, and we evaluate this design's performance using ray tracing simulations in Ansys Zemax OpticStudio non-sequential mode. In Section~\ref{sec:telescope_FEM_PS}, we discuss VIPER's front-end module and combined pupil slicer and double scrambler.
\section{VIPA Spectrograph Overview}\label{sec:VIPA_spec_overview}

\subsection{Virtually Imaged Phased Array (VIPA)}

\begin{figure}[H]
\centering\includegraphics[width=0.6\textwidth]{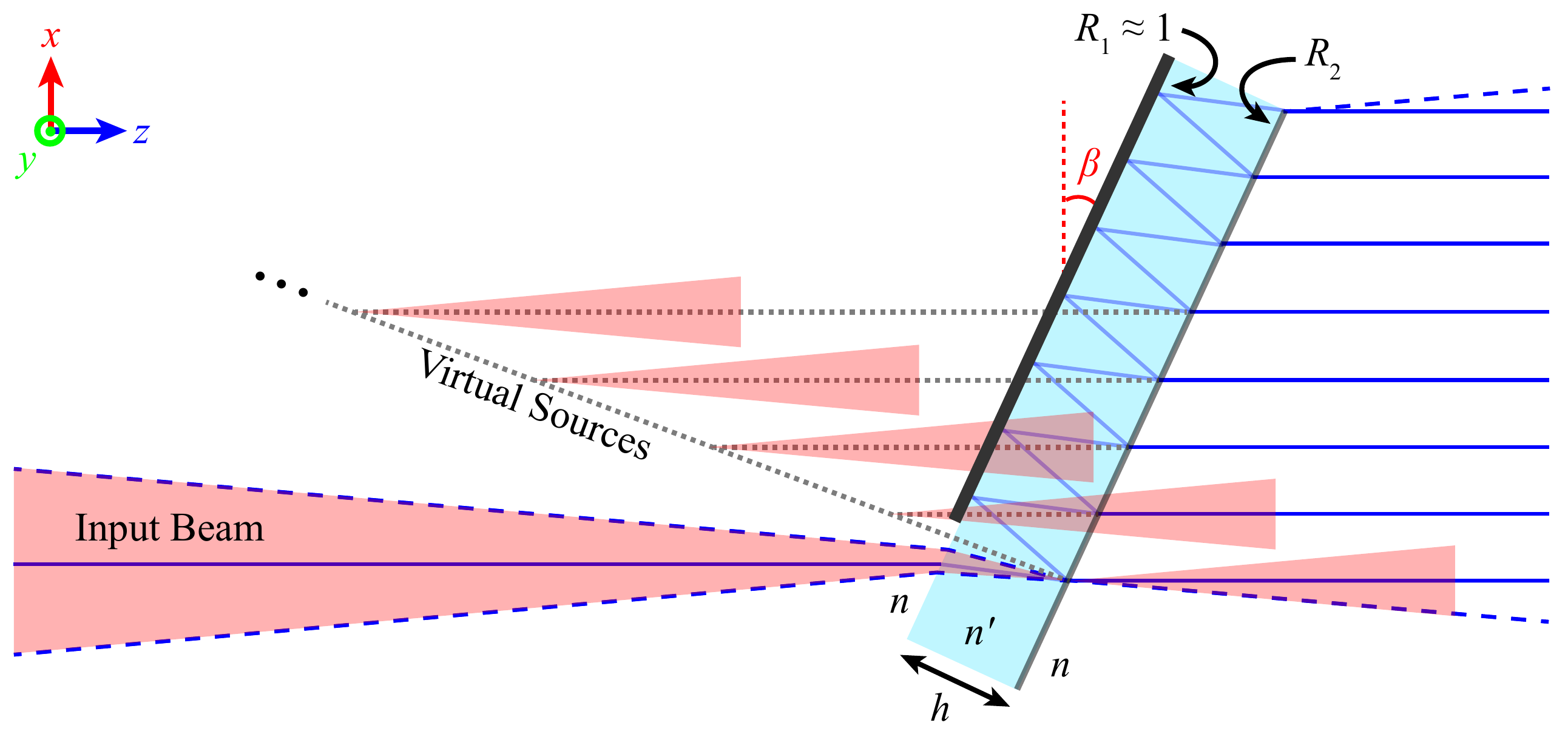}
\caption{Illustration of a Virtually Imaged Phased Array (VIPA). Multiple reflections between the two internally reflective surfaces create a phased array of virtual sources that interfere to achieve high spectral resolution.}\label{fig:VIPA}
\end{figure}

Figure~\ref{fig:VIPA} shows an illustration of a VIPA. A VIPA is basically a modified Fabry-Perot etalon with a transmissive entrance window. It is a glass slab with two internally partially-reflective surfaces. Light reflects between these two surfaces, accumulating an optical phase difference in each round trip. When the round-trip phase difference is an integer multiple of $2\pi$, a resonance condition is met and there is an intensity peak. Since this round-trip phase difference is wavelength-dependent, there is dispersion. In Figure~\ref{fig:VIPA}, the VIPA disperses light along the $x$-direction.

\begin{figure}[H]
    \centering
    \includegraphics[width=0.8\textwidth]{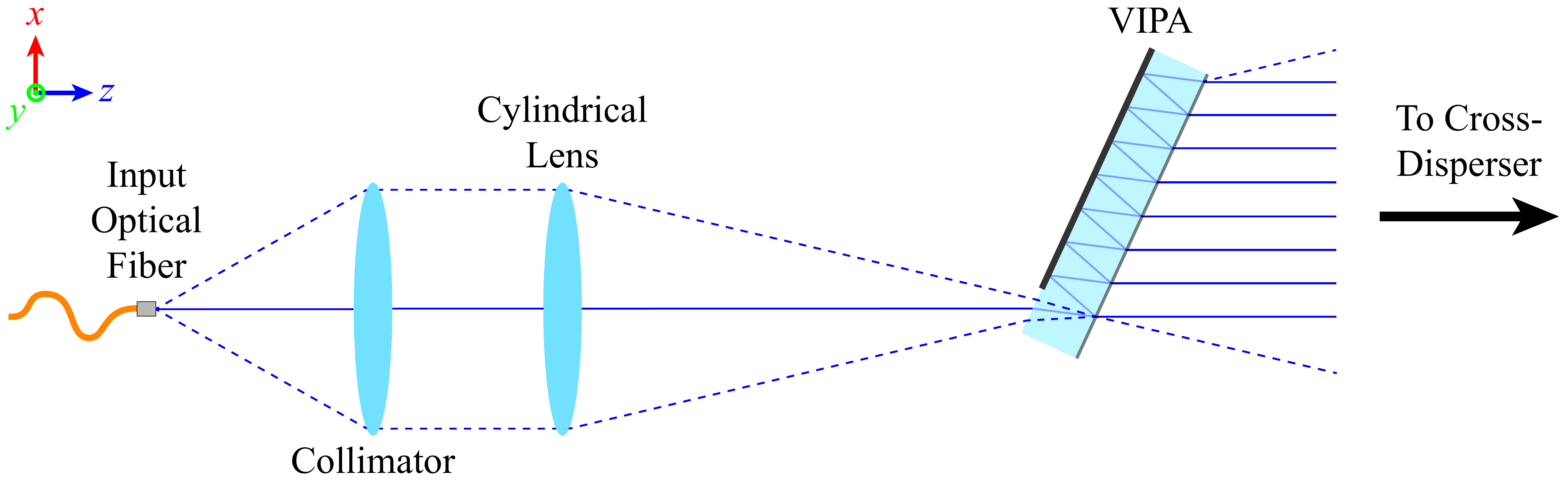}
    \caption{Optical system used to couple light from an optical fiber into a VIPA. This is also a side view of the cross-dispersed VIPA spectrograph shown in Figure \ref{fig:VIPA_spec_top}, from the input fiber to the VIPA.}
    \label{fig:VIPA_spec_side}
\end{figure}

Figure~\ref{fig:VIPA_spec_side} shows an illustration of a typical optical system used to couple light from an optical fiber into a VIPA. Light comes out of the input optical fiber and is collimated by a collimator. The collimated beam is then line-focused onto the back surface of the VIPA using a cylindrical lens. This cylindrical lens has power only along the VIPA dispersion direction ($x$-direction in this case).

\begin{figure}[H]
\centering\includegraphics[width=0.5\textwidth]{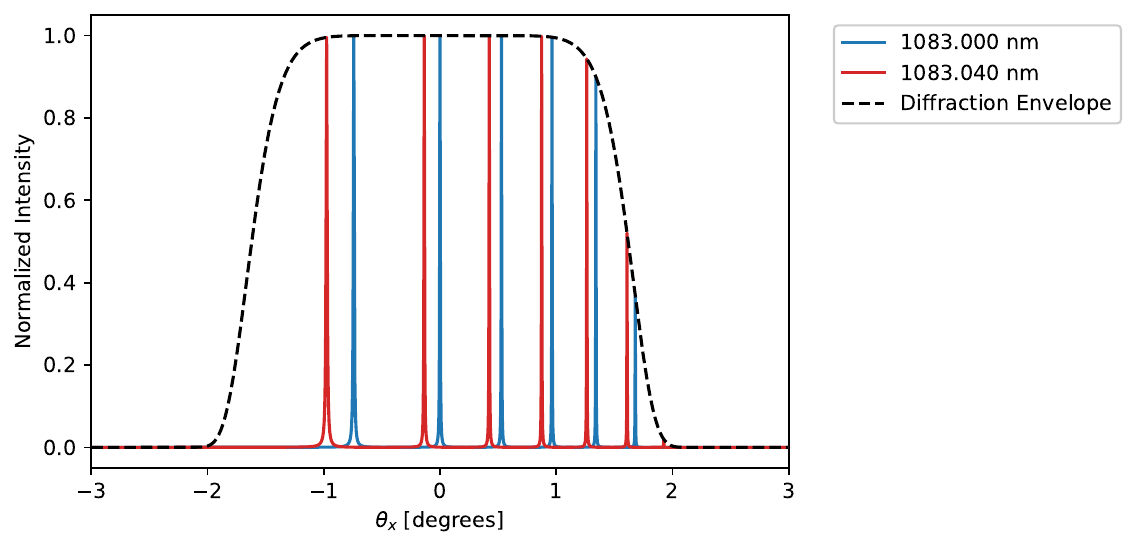}
\caption{Example output angular spectrum of a VIPA at two wavelengths. The intensity peaks corresponding to different wavelengths are at different output angles. Each intensity peak is Lorentzian, as in a Fabry-Perot etalon. Each intensity peak corresponds to one VIPA diffraction order. The output angular spectrum is modulated by the black dashed curve, called the ``diffraction envelope'', which is a scaled version of the far-field of the input optical fiber.}\label{fig:VIPA_output_ang_spec}
\end{figure}

Figure~\ref{fig:VIPA_output_ang_spec} shows an example of the output angular spectrum of the VIPA in the optical system shown in Figure~\ref{fig:VIPA_spec_side}. We can see that the intensity peaks of different wavelengths are at different output angles, due to the wavelength-dependence of the round-trip optical phase difference. Each intensity peak is Lorentzian, as in a Fabry-Perot etalon, and corresponds to one diffraction order of the VIPA. The output angular spectrum is modulated by the black dashed curve, which is sometimes called the ``diffraction envelope'' in the literature\cite{Zhu2023}. This diffraction envelope is a scaled version of the angular spectrum of the input source (input fiber), scaled by the collimator and the cylindrical lens, and compressed along the VIPA dispersion direction. In other words, this diffraction envelope is a scaled version of the input optical fiber's far-field. This diffraction envelope is analogous to the blaze function in a cross-dispersed echelle spectrograph, and is imprinted upon each order.

%%%%%%%%%%%%%%%%%%%%%%%%%%%%%%%%%%%%%%%%%%%%%%%%%%%%%%%%%%%%%%%%%%%%
%%%%%%%%%%%%%%%%%%%%%%%%%%%%%%%%%%%%%%%%%%%%%%%%%%%%%%%%%%%%%%%%%%%%

\subsection{Cross-dispersed VIPA spectrograph}

\begin{figure}[H]
    \centering
    \includegraphics[width=0.75\textwidth]{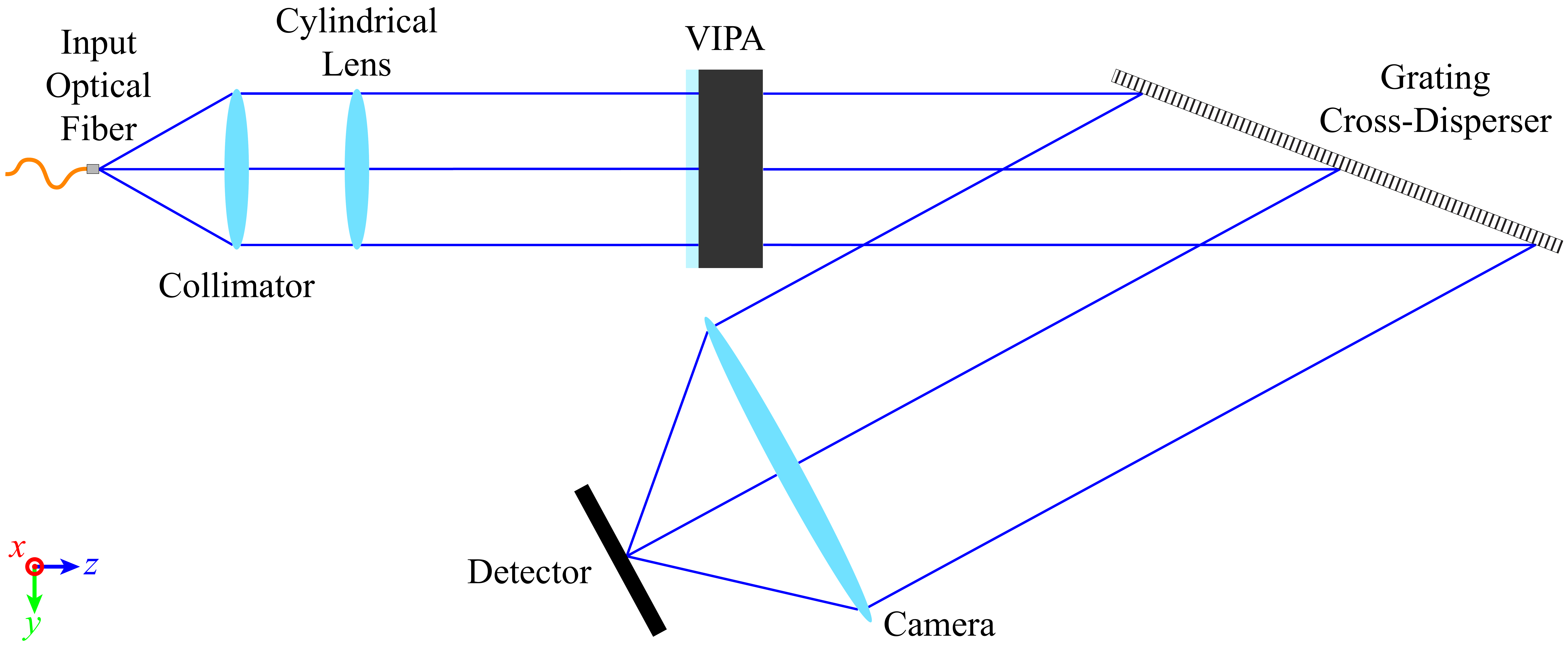}
    \caption{Illustration of a cross-dispersed VIPA spectrograph. Light from the input optical fiber is collimated by a collimator and is then line-focused by a cylindrical lens onto the back surface of the VIPA. The VIPA disperses light in the $x$-direction. A grating then cross disperses the beam in the direction orthogonal to the VIPA dispersion direction. Finally, a camera focuses the spectrum onto a detector.}
    \label{fig:VIPA_spec_top}
\end{figure}

A VIPA spectrograph is similar to a grating spectrograph, in that both have an input optical fiber (slit), a collimator, some camera optics, and a detector. However, in a VIPA spectrograph, in place of the diffraction grating in the grating spectrograph, we have a cylindrical lens and a VIPA. Figure~\ref{fig:VIPA_spec_top} shows a top view of a cross-dispersed VIPA spectrograph. The first part of the the spectrograph consists of the input optical fiber, the collimator, the cylindrical lens, and the VIPA (as seen in Figure~\ref{fig:VIPA_spec_side}, which is also a side view of Figure~\ref{fig:VIPA_spec_top} from the input fiber to the VIPA). Since the VIPA has a narrow free spectral range, on the order of 0.1~nm to 0.2~nm, the spectrum must be cross-dispersed after the VIPA in order to separate the VIPA diffraction orders. This is analogous to cross-dispersion in an echelle spectrograph. Hence, following the VIPA, a diffraction grating is placed to disperse light in the direction orthogonal to the VIPA dispersion direction. Finally, some camera optics focus the spectrum onto a detector. Notice, in Figure~\ref{fig:VIPA_spec_top}, that between the collimator and the camera, the beam remains collimated along the cross-dispersion direction.

%%%%%%%%%%%%%%%%%%%%%%%%%%%%%%%%%%%%%%%%%%%%%%%%%%%%%%%%%%%%%%%%%%%%
%%%%%%%%%%%%%%%%%%%%%%%%%%%%%%%%%%%%%%%%%%%%%%%%%%%%%%%%%%%%%%%%%%%%

\subsection{Design challenges of a multimode fiber-fed VIPA spectrograph}\label{sec:MMF_VIPA_spec_challenges}

Most VIPA spectrographs in the literature are fed by single-mode optical fiber rather than multimode optical fiber. This is because a single-mode fiber has a more stable beam profile, a well-defined spatial and angular extent that can be approximated by a Gaussian beam\cite{SalehTeich}, and a smaller etendue\cite{Bourdarot2018}, which generally makes optical design easier. However, due to atmospheric seeing, a single-mode fiber-fed astronomical VIPA spectrograph requires the use of adaptive optics so that enough light can be injected into the small single-mode fiber at the telescope\cite{Bourdarot2018,Carlotti2022}. A larger multimode fiber feed would relax this requirement, allowing VIPA spectrographs to operate on telescopes without adaptive optics, enabling higher instrument throughput and broadening their accessibility. However, the use of a multimode fiber feed instead of a single-mode fiber feed gives rise to several challenges:
\begin{enumerate}
    \item \textbf{More cross-dispersion required:} The cross-dispersion direction is still a conventional grating spectrograph (see Figure~\ref{fig:VIPA_spec_top}). Hence, if the input fiber diameter is larger, then the fiber's image at the detector is also larger, and so there needs to be more separation between the cross-dispersed VIPA diffraction orders. Since the VIPA's clear aperture limits the collimator's focal length\cite{Leung2025}, in order to adequately separate the orders, a cross-disperser with a very high angular dispersion is required. It may be difficult to obtain such a cross-disperser with high enough diffraction efficiency. One alternative solution is to use an additional slit to decrease the fiber size along the cross-dispersion direction\cite{Zhu2023}. However, this discards a significant amount of light.
    \item \textbf{Fiber far-field imprinted on diffraction envelope:} Since the diffraction envelope is a scaled version of the input fiber's far-field, any variations in fiber illumination (e.g., from seeing) will be imprinted upon every order, across the spectrum. This is very problematic for astronomical applications. Compared to a multimode fiber feed, a single-mode fiber feed is less susceptible to this effect because it can support only a nearly Gaussian far-field.
    \item \textbf{VIPA coupling loss:} There is a limit to how much light can be coupled into a VIPA. A VIPA in a particular configuration can only accept a certain amount of etendue\cite{Leung2026}. This is more difficult with a multimode fiber, which has a larger etendue compared to a single-mode fiber.
\end{enumerate}
Our design for VIPER, presented in Sections \ref{sec:inst_overview} and \ref{sec:optical_design}, addresses all three of these challenges.
\section{VIPER instrument overview}\label{sec:inst_overview}

\subsection{Science and instrument requirements}

VIPER is specifically designed to observe the metastable helium triplet line at 1083~nm. This helium line is of particular importance in several areas of astrophysics, as it is a tracer for exoplanet atmospheric escape\cite{Oklopi2018}, stellar activity\cite{Strader2015}, stellar mass loss\cite{Dupree1986}, and galactic chemical evolution\cite{Cooke2022}. VIPER's primary science goal is to detect anisotropic atmospheric escape from gaseous exoplanets, which is traced by the helium 1083~nm line. Since the line profiles can be kinematically broadened rather than thermally broadened \cite{Nail2025}, high-resolution spectra can reveal more details about the kinematics of the escaping gas, providing constraints on the underlying physical mechanisms driving atmospheric escape. Since atmospheric escape is believed\cite{Owen2022} to be an important contributor to observed exoplanet demographic features like the radius gap\cite{Fulton2017} and the Neptune desert\cite{Mazeh2016}, better understanding atmospheric escape will ultimately help us better constrain models of planet formation, evolution, and migration.

A more detailed description of VIPER's science drivers can be found in Ref.~\citenum{Leung2025}. In that work, we derived VIPER's instrument requirements as informed from the science. For the atmospheric escape science case, our requirements were a bandpass of $\geq 3.54$~nm about $1083$~nm, a resolving power of 300,000, and an instrment efficiency of around $\gtrsim 30\%$, with a limiting $J$-band apparent magnitude of 10. VIPER needs to be compatible with the 100 \textmu m diameter circular optical fiber feed (f/6) at the Tillinghast Telescope.

\subsection{Instrument overview}

\begin{figure}[H]
\centering\includegraphics[width=\textwidth]{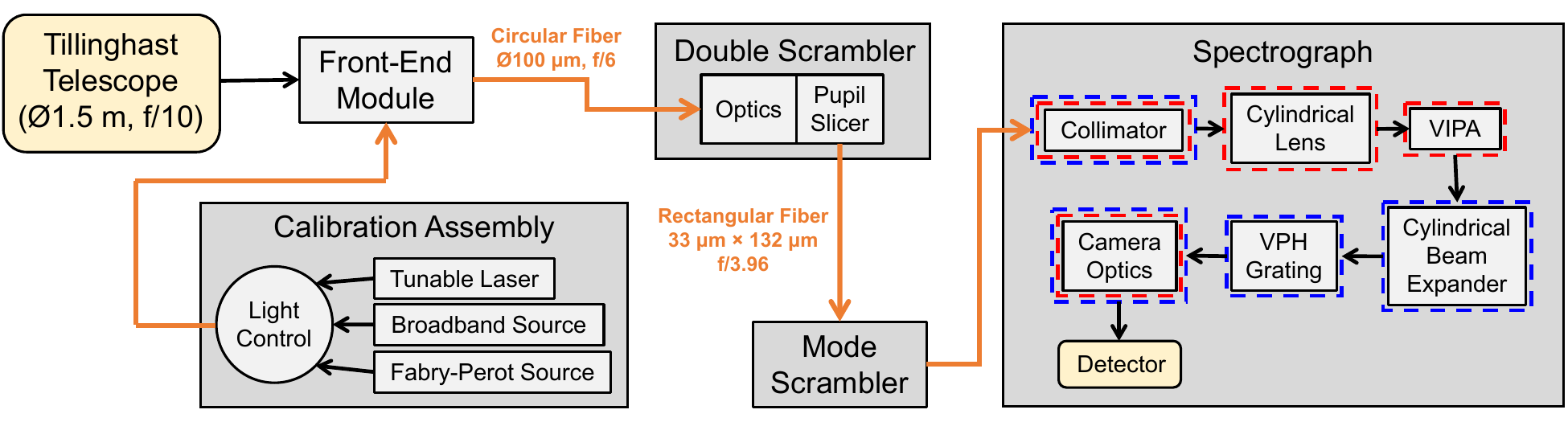}
\caption{VIPER instrument architecture. There are several distinct optomechanical subsystems: the front-end module, the calibration assembly, the double scrambler, the mode scrambler, and the spectrograph. Orange arrows denote connections via optical fiber. In the spectrograph box, elements with a red dashed border are those that affect light in the VIPA dispersion direction, while elements with a blue dashed border are those that affect light in the cross-dispersion direction.}\label{fig:VIPER_inst_arch}
\end{figure}

Figure~\ref{fig:VIPER_inst_arch} shows a schematic of the VIPER instrument architecture. Light comes into the Tillinghast Telescope, which has a diameter of 1.5~m and a focal ratio of f/10, and then enters a front-end module. This front-end module comprises a guiding camera, a focal reducer, and a tip-tilt mirror that feeds light into a Ø100~\textmu m circular optical fiber at f/6. This front-end module is currently used by the Tillinghast Reflector Echelle Spectrograph (TRES\cite{Szentgyorgyi2007}), and was upgraded in 2021. The front-end module is described in more detail in Section~\ref{sec:telescope_FEM}. The front end-module is also used to inject calibration light from the calibration assembly. The calibration assembly comprises several calibration sources: a tunable laser, a broadband source, and a Fabry-Perot source.

The Ø100~\textmu m circular fiber is then routed to the double scrambler, which is described in more detail in Section~\ref{sec:slicer}. This double scrambler is a pupil slicer. It slices and reimages the far-field of the Ø100~\textmu m circular fiber at f/6 onto a 33~\textmu m~$\times$~132~\textmu m rectangular fiber at f/3.96. Etendue is conserved in this process. This rectangular fiber is then agitated using a mode scrambler\cite{Leung2025MS} to mitigate the phenomenon of modal noise\cite{Epworth1979}, and is then fed to the spectrograph. The rectangular fiber is attached to the spectrograph such that the shorter end of the fiber is parallel to the VIPA dispersion direction. The optical design of the spectrograph is discussed in detail in Section~\ref{sec:optical_design}.

%%%%%%%%%%%%%%%%%%%%%%%%%%%%%%%%%%%%%%%%%%%%%%%%%%%%%%%%%%%%%%%%%%%%
%%%%%%%%%%%%%%%%%%%%%%%%%%%%%%%%%%%%%%%%%%%%%%%%%%%%%%%%%%%%%%%%%%%%

\section{VIPER Optical Design}\label{sec:optical_design}

\subsection{Optical design overview}

\begin{figure}[H]
\centering\includegraphics[width=\textwidth]{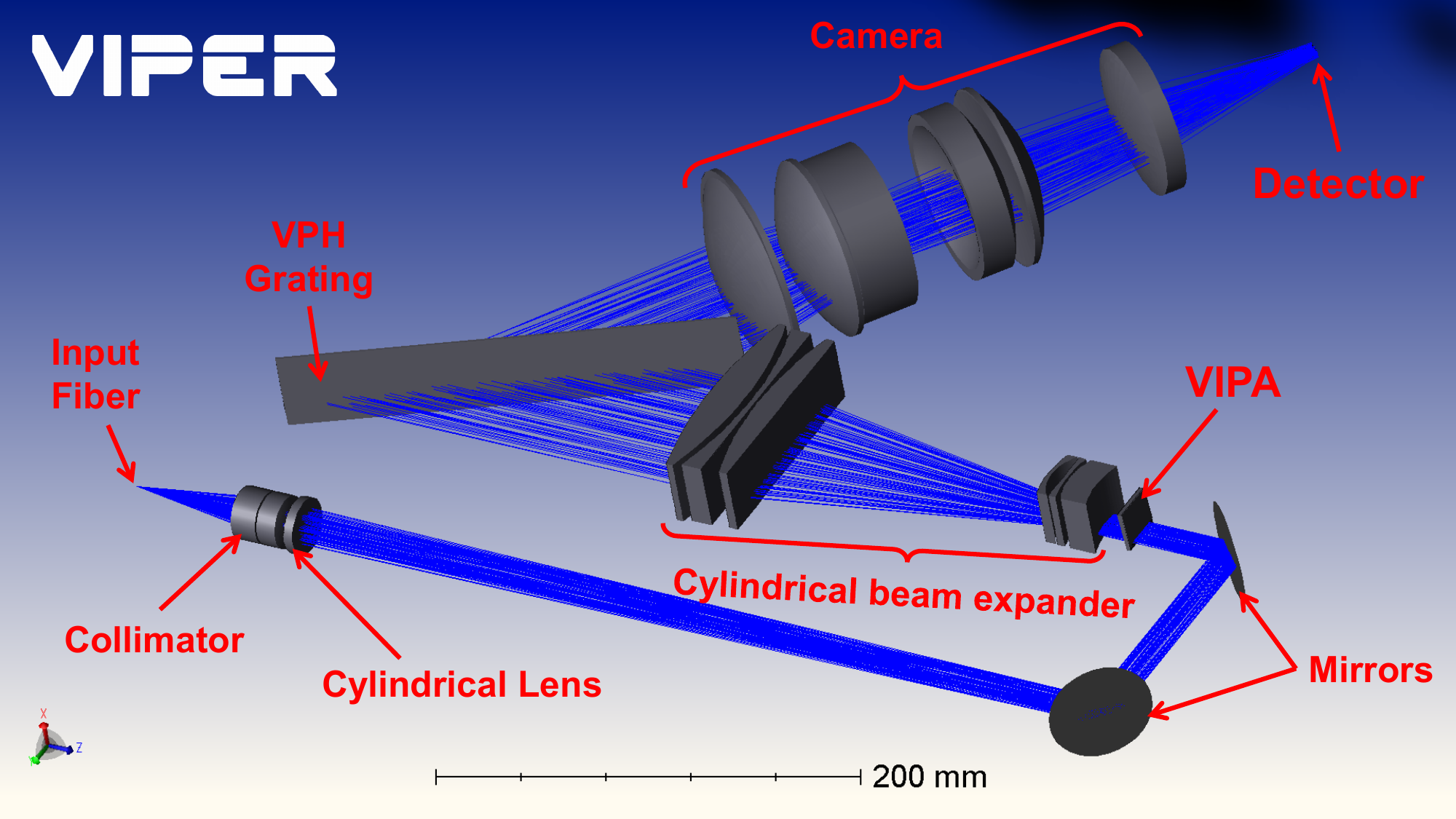}
\caption{3D shaded model of VIPER's optical design, from Zemax OpticStudio non-sequential mode.}\label{fig:VIPER_Shaded}
\end{figure}

\begin{figure}[H]
\centering\includegraphics[width=0.6\textwidth]{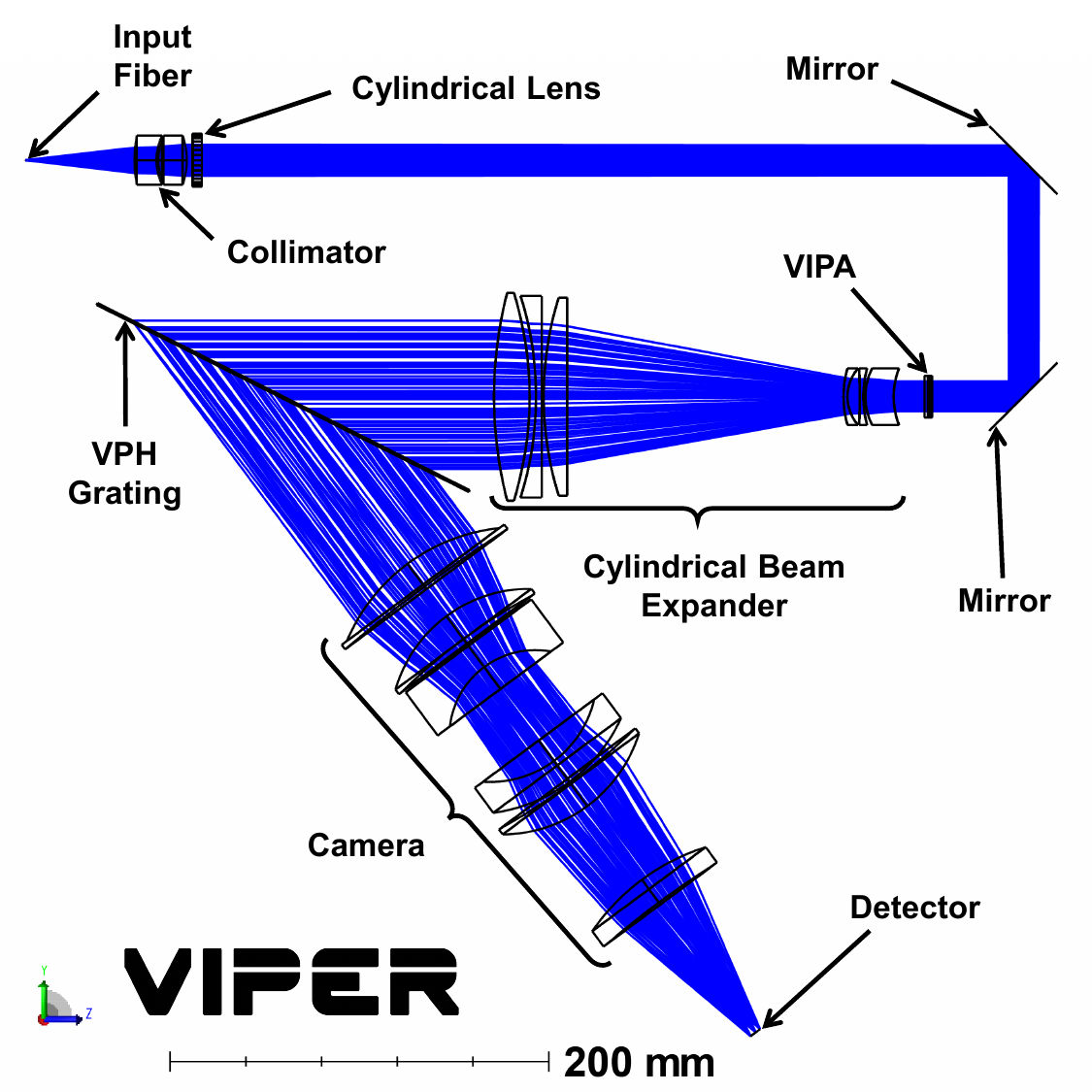}
\caption{Optical design of VIPER, from Zemax OpticStudio non-sequential mode. The entire design fits inside a square that is roughly 50~cm by 50~cm.}\label{fig:VIPER_3D}
\end{figure}

Figures~\ref{fig:VIPER_Shaded} and \ref{fig:VIPER_3D} show the optical design of VIPER. Light comes out of the input rectangular optical fiber and is collimated by a collimator. The collimated beam is then line-focused by a cylindrical lens on to the back surface of the VIPA. The cylindrical lens has a long focal length, and so the converging beam is folded using two dielectric mirrors. Recall that the cylindrical lens has power only along the VIPA dispersion direction. 

Following the VIPA is a $5\times$ cylindrical Galilean beam expander. Note that after the VIPA, the beam is still collimated along the cross-dispersion direction, and this cylindrical beam expander has power only along the cross-dispersion direction. This cylindrical beam expander effectively increases the focal length of the collimator along the cross-dispersion direction, mitigating the first challenge discussed in Section~\ref{sec:MMF_VIPA_spec_challenges}.

After the cylindrical beam expander, a Volume Phase Holographic (VPH) grating cross-disperses the light in the direction orthogonal to the VIPA dispersion direction. Finally, some camera optics focus the spectrum onto a detector. As seen in Figure~\ref{fig:VIPER_3D}, the entire optical design fits in a roughly 50~cm by 50~cm square. Table~\ref{tab:design} shows a summary of VIPER's design parameters. These parameters were selected following a design procedure developed in our previous work\cite{Leung2025}, with some modifications that will be discussed in a future work. The next subsections will discuss each component of VIPER in detail.

\begin{table}[htb]
\centering
\begin{tabular}{ | m{0.45\textwidth} | m{0.1\textwidth}| m{0.2\textwidth} | } 
  \hline
  Variable & Symbol & Value \\ 
  \hline
  \hline
  Design resolving power & $\mathcal{R}_\mathrm{des}$ & 300,000 \\
  Design wavelength & $\lambda_\mathrm{des}$ & 1083 nm \\
  Input fiber $x$-width & $s_x$ & $33$ \textmu m \\
  Input fiber $y$-width & $s_y$ & $132$ \textmu m \\
  Input fiber output focal ratio & $F$ & f/3.96 \\
  \hline
  VIPA thickness & $h$ & $2.20$ mm \\
  VIPA tilt angle & $\beta$ & $4.750^\circ$ \\
  Refractive index of VIPA at $\lambda_\mathrm{des}$ & $n'$ & $1.4494$ (fused silica) \\
  Refractive index of surrounding medium & $n$ & $1.0003$ (air) \\
  VIPA front surface internal reflectivity & $R_1$ & $99.5\%$ \\
  VIPA back surface internal reflectivity & $R_2$ & $95\%$ \\
  VIPA clear aperture & $D_\mathrm{VIPA}$ & $18$ mm\\
  \hline
  Collimator focal length & $f_\mathrm{coll}$ & $71.28$ mm \\
  Cylindrical lens focal length & $f_c$ & $612.00$ mm \\
  Camera focal length & $f_\mathrm{cam}$ & $175.70$ mm \\
  Collimator focal ratio & $F_\mathrm{coll}$ & f/3.96 \\
  Cylindrical lens focal ratio & $F_c$ & f/34 \\
  Camera design focal ratio & $F_\mathrm{cam}$ & f/1.757\\
  \hline
  Cylindrical beam expander L1 focal length & $f_\mathrm{L1}$ & $-44.00$ mm \\
  Cylindrical beam expander L2 focal length & $f_\mathrm{L1}$ & $220.00$ mm \\
  Cylindrical beam expander L1 design focal ratio & $F_\mathrm{L1}$ & f/2 \\
  Cylindrical beam expander L2 design focal ratio & $F_\mathrm{L2}$ & f/2 \\
  \hline
  Cross-disperser grating line density & $G$ & $1650$ lines/mm \\
  Cross-disperser grating incidence angle & $\alpha_0$ & $63.3^\circ$ \\
  Cross-disperser grating diffraction angle at $\lambda_\mathrm{des}$ & $\beta_0$ & $63.258^\circ$ \\
  \hline
\end{tabular}
\caption{Summary of VIPER's design parameters.}\label{tab:design}
\end{table}

%%%%%%%%%%%%%%%%%%%%%%%%%%%%%%%%%%%%%%%%%%%%%%%%%%%%%%%%%%%%%%%%%%%%
%%%%%%%%%%%%%%%%%%%%%%%%%%%%%%%%%%%%%%%%%%%%%%%%%%%%%%%%%%%%%%%%%%%%

\subsection{Collimator and cylindrical lens}

\begin{figure}[H]
\centering\includegraphics[width=0.7\textwidth]{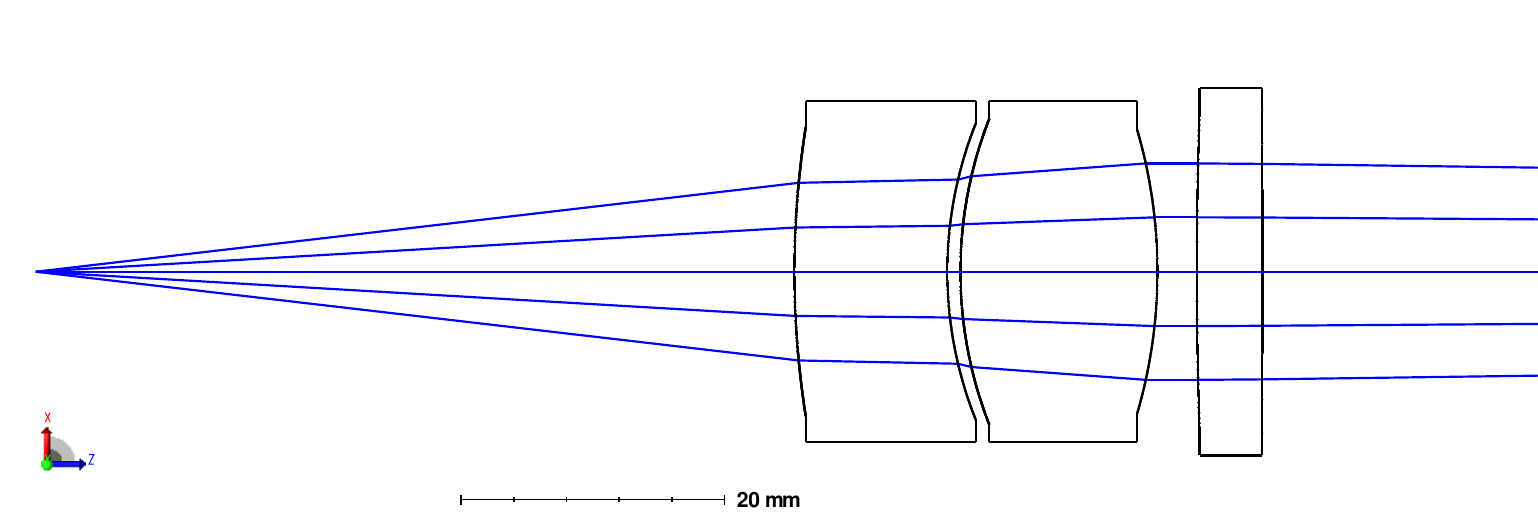}
\caption{Collimator and cylindrical lens used in VIPER. The collimator is the air-spaced doublet on the left, and the cylindrical lens is the singlet on the right.}\label{fig:Zemax_coll_cyl}
\end{figure}

\begin{figure}[H]
    \centering
    \begin{subfigure}{0.49\textwidth}
        \centering
        \includegraphics[width=\textwidth]{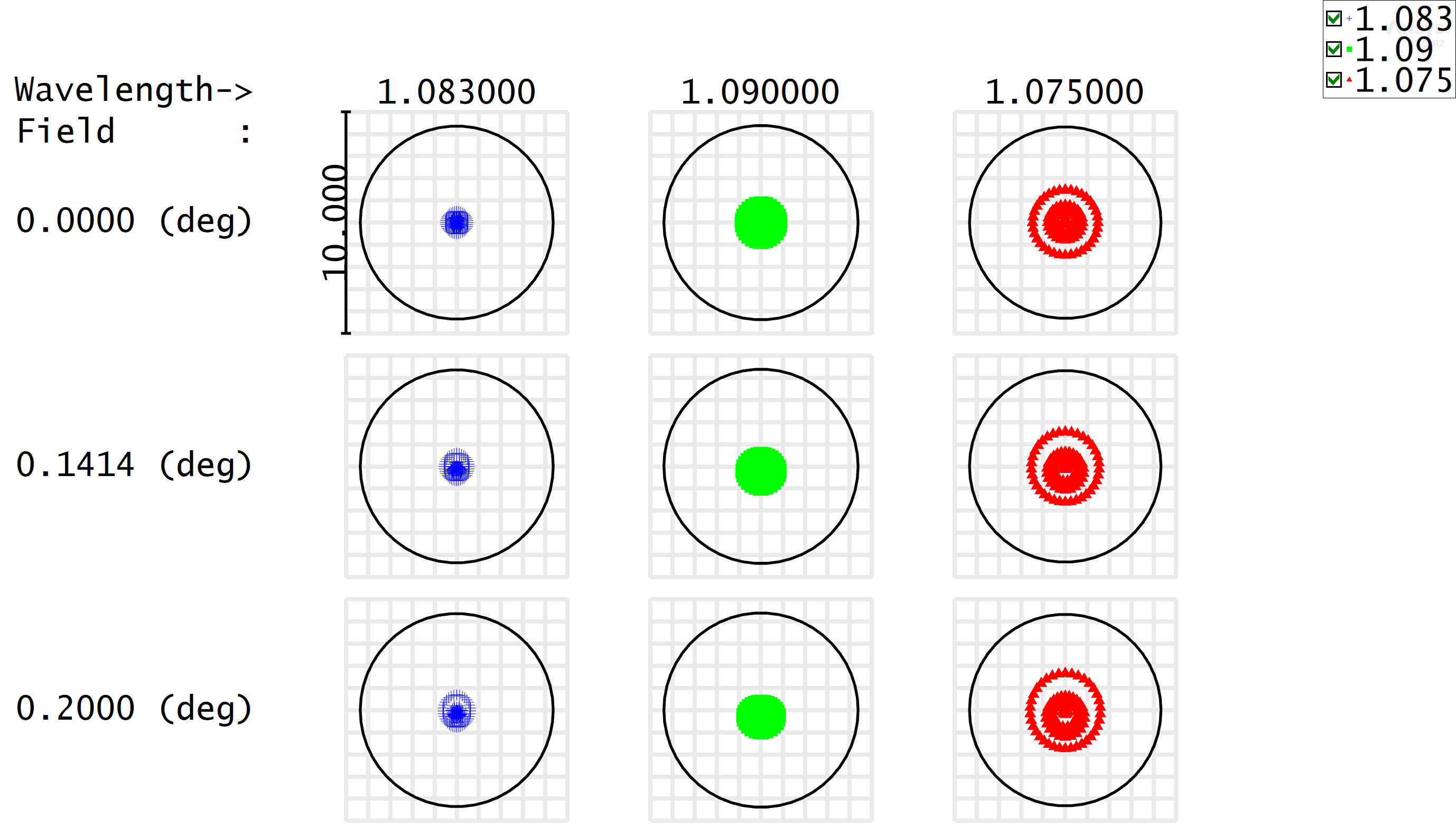}
        \caption{Spot diagram for collimator.}
        \label{fig:Collimator_MatrixSpotDiagram}
    \end{subfigure}
    \begin{subfigure}{0.49\textwidth}
        \centering
        \includegraphics[width=\textwidth]{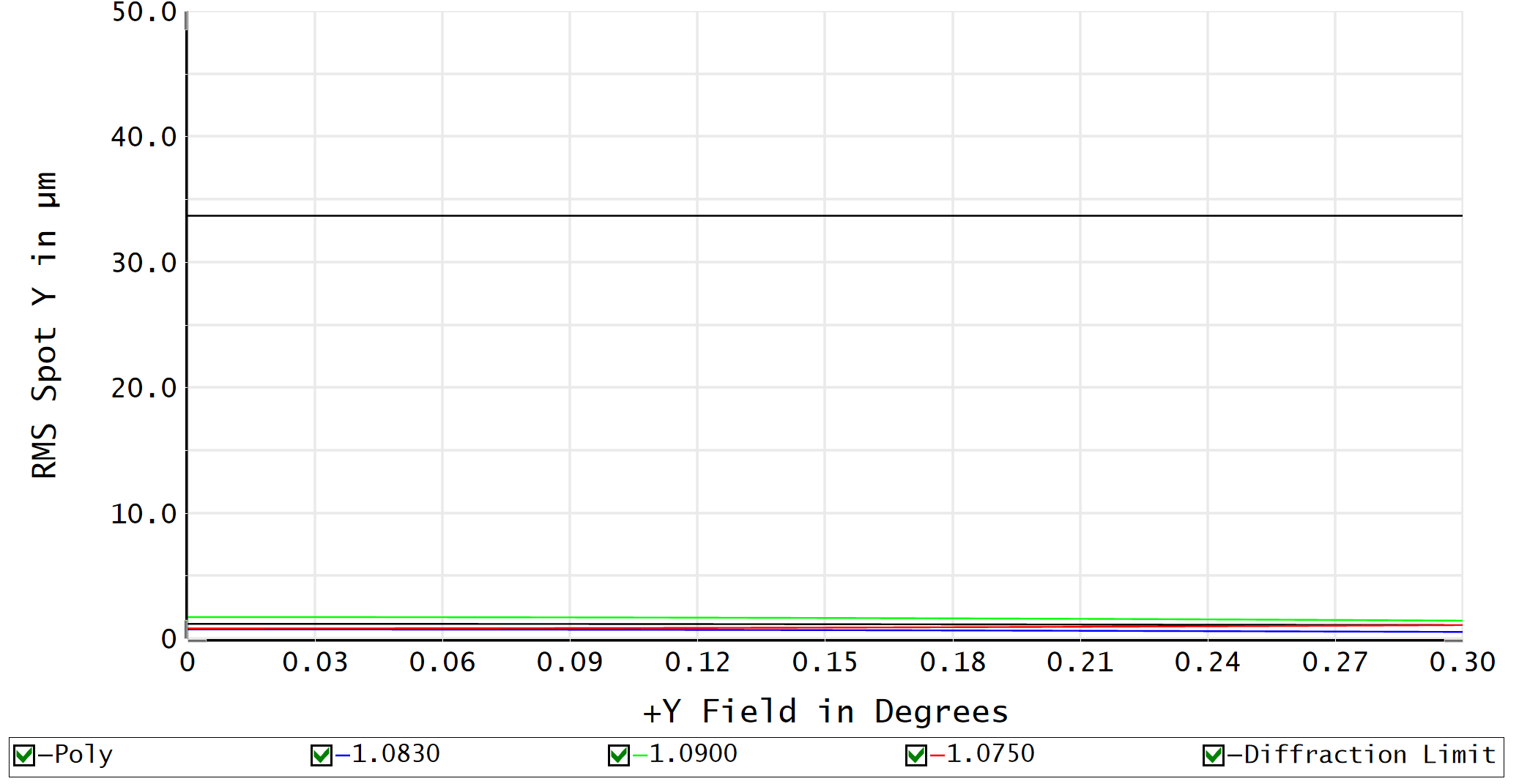}
        \caption{RMS spot VS field for cylindrical lens.}
        \label{fig:CylindricalLens_RMSvsField}
    \end{subfigure}
    \caption{Metrics for the collimator and cylindrical lens. Wavelengths are in \textmu m. In the spot diagram for the collimator, the scale bar denotes 10 \textmu m, and the black circles denote the Airy radius.}
    \label{fig:cyl_col_metrics}
\end{figure}

Figure~\ref{fig:Zemax_coll_cyl} shows an illustration of the collimator and cylindrical lens used in VIPER. The collimator is an air-spaced doublet, consisting of two spherical lenses. It has a focal length of $f_\mathrm{coll}=71.28$ mm and is matched the output focal ratio of the input rectangular fiber, which is f/3.96. The collimated beam has a diameter of 18~mm, which is the clear aperture of the VIPA used in our design. The collimator has diffraction-limited performance, as seen in Figure~\ref{fig:Collimator_MatrixSpotDiagram}. The cylindrical lens is a singlet, with a focal length of $f_c=612.00$ mm and a focal ratio of f/34. The cylindrical lens has diffraction-limited performance, as seen in Figure~\ref{fig:CylindricalLens_RMSvsField}.

%%%%%%%%%%%%%%%%%%%%%%%%%%%%%%%%%%%%%%%%%%%%%%%%%%%%%%%%%%%%%%%%%%%%

\subsection{VIPA}

\begin{figure}[H]
\centering\includegraphics[width=0.85\textwidth]{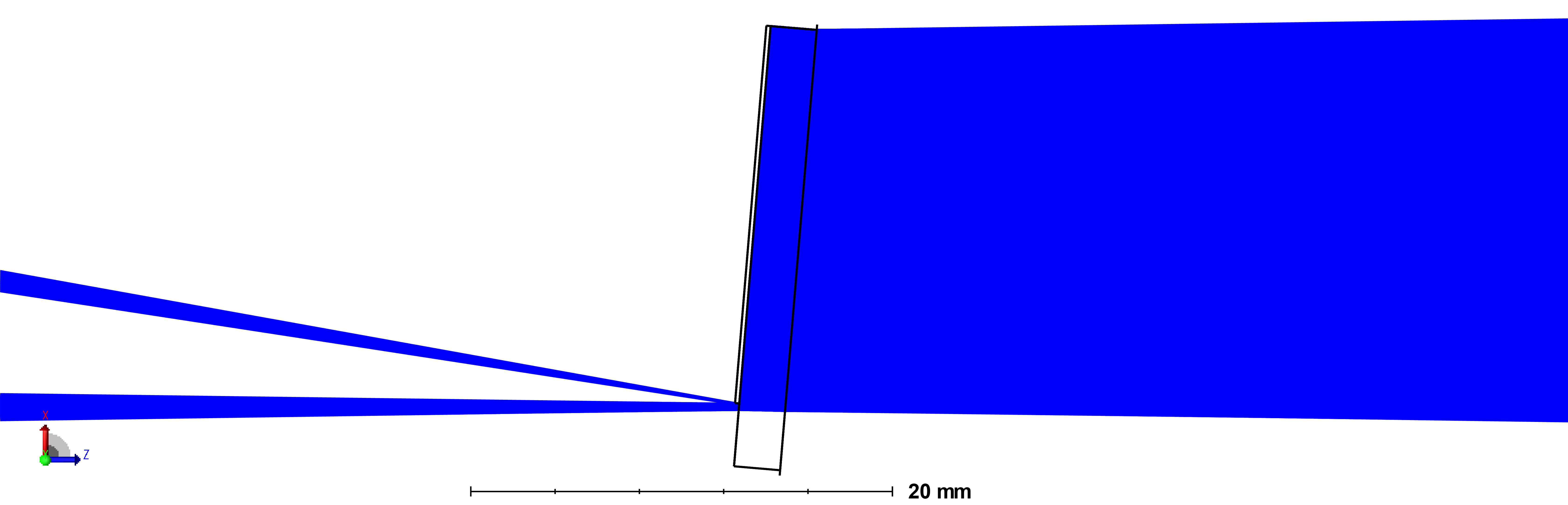}
\caption{VIPA used in VIPER. A portion (around $14\%$) of the light coupled into the VIPA is lost by escaping back through the entrance window.}\label{fig:Zemax_VIPA}
\end{figure}

Figure~\ref{fig:Zemax_VIPA} shows a model of the VIPA used in VIPER, from Zemax OpticStudio non-sequential mode. The VIPA is from LightMachinery Inc., and is made of fused silica, with a thickness of $h=2.20$~mm and a clear aperture of $D_\mathrm{VIPA}=18$~mm. The reflectivities of the front and back surfaces are $R_1=99.5\%$ and $R_2=95\%$ respectively. In our design, we selected a VIPA tilt angle of $4.750^\circ$. This value was specifically selected to maximize the total transmission of the instrument, considering the VIPA coupling loss\cite{Leung2026} and some other effects that will be discussed in a future work. As seen in Figure~\ref{fig:Zemax_VIPA}, there is some VIPA coupling loss, as some of the light coupled into the VIPA is lost by escaping back through the entrance window; this is by design. For a comprehensive discussion on VIPA coupling loss and transmission, see Ref.~\citenum{Leung2026}.

%%%%%%%%%%%%%%%%%%%%%%%%%%%%%%%%%%%%%%%%%%%%%%%%%%%%%%%%%%%%%%%%%%%%

\subsection{Cylindrical beam expander}

\begin{figure}[H]
\centering\includegraphics[width=0.65\textwidth]{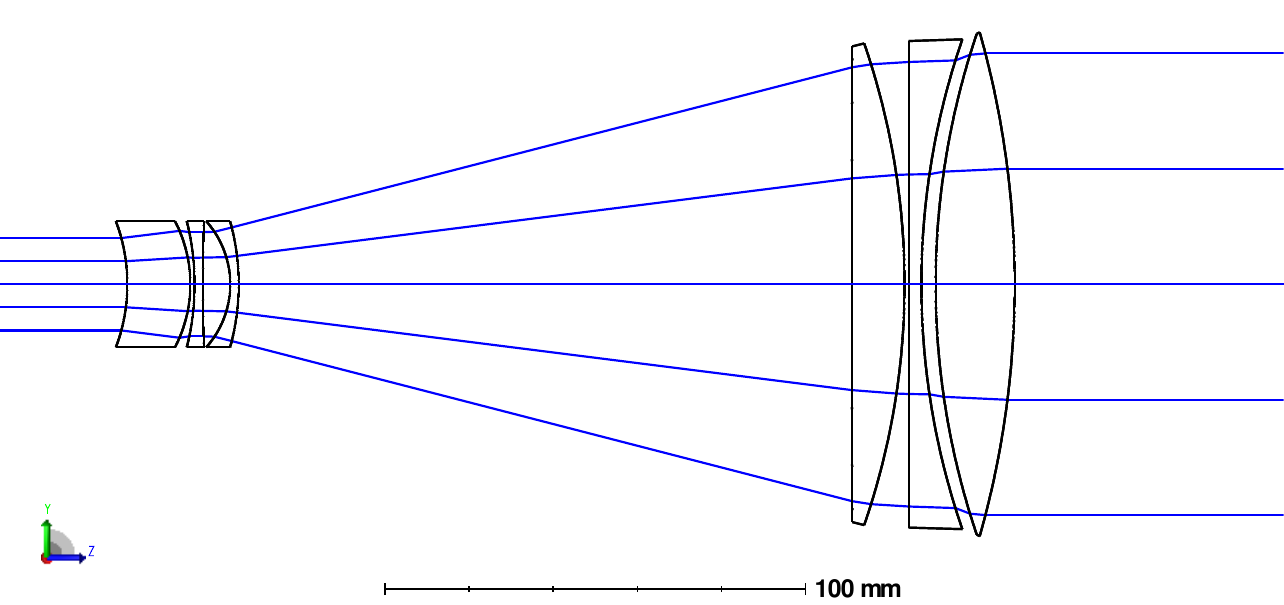}
\caption{Cylindrical Galilean $5\times$ beam expander used in VIPER, after the VIPA. This beam expander consists of two lens groups, L1 (diverging) and L2 (converging), which are air-spaced triplets and are designed with a focal ratio of f/2. The cylindrical elements have power along the cross-dispersion direction.}\label{fig:Zemax_BeamExpander}
\end{figure}

\begin{figure}[H]
    \centering
    \begin{subfigure}{0.49\textwidth}
        \centering
        \includegraphics[width=\textwidth]{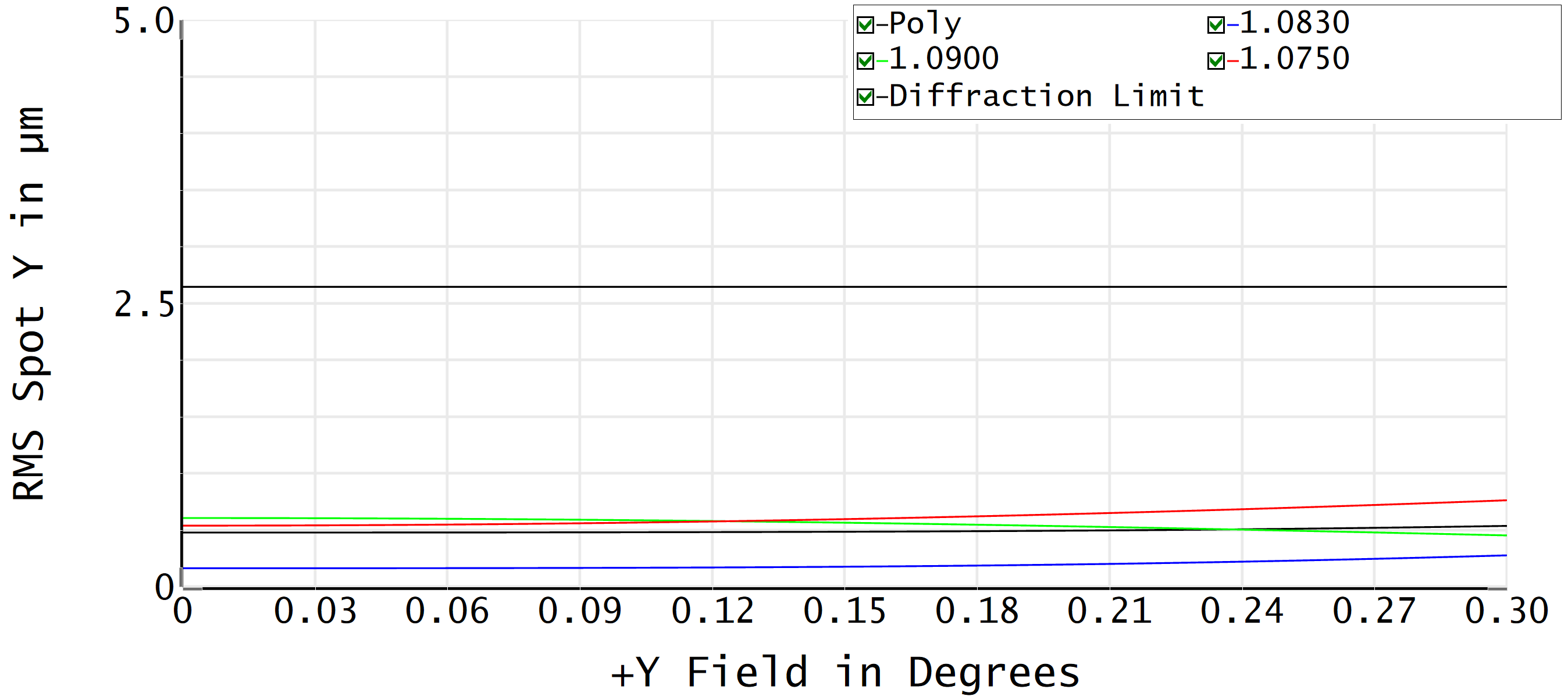}
        \caption{RMS spot VS field for L1.}
        \label{fig:L1_RMSvsField}
    \end{subfigure}
    \begin{subfigure}{0.49\textwidth}
        \centering
        \includegraphics[width=\textwidth]{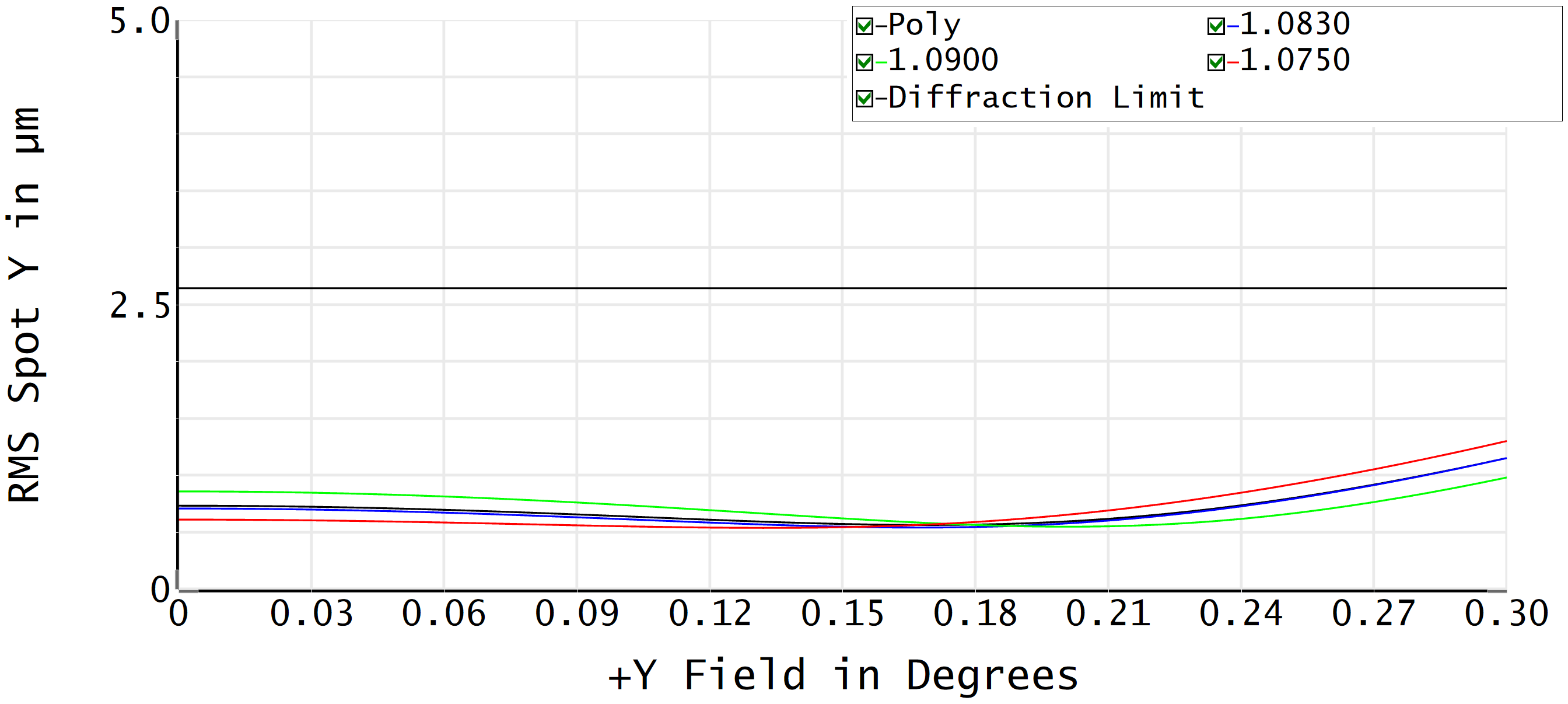}
        \caption{RMS spot VS field for L2.}
        \label{fig:L2_RMSvsField}
    \end{subfigure}
    \caption{RMS spot VS field for L1 and L2 in the cylindrical Galilean beam expander, at different wavelengths (in \textmu m).}
    \label{fig:beam_expander_metrics}
\end{figure}

Figure \ref{fig:Zemax_BeamExpander} shows the $5\times$ cylindrical Galilean beam expander used in VIPER. We used a Galilean beam expander instead of a Keplerian beam expander so that our design is more compact. The cylindrical beam expander consists of two lens groups, L1 and L2. L1 is diverging and has a focal length of $f_\mathrm{L1}=-44.00$~mm. L2 is converging and has a focal length of $f_\mathrm{L2}=220.00$~mm. L1 and L2 are both air-spaced triplets, and were designed with a focal ratio of f/2. We selected this relatively fast focal ratio so that the beam expander can be compact, at the expense of using more lenses to correct for aberrations. The cylindrical elements have power only along the cross-dispersion direction. L1 has diffraction-limited performance and L2 has near-diffraction-limited performance, as seen in Figures~\ref{fig:L1_RMSvsField} and \ref{fig:L2_RMSvsField} respectively. After the cylindrical beam expander, the 18~mm wide collimated beam in the cross-dispersion direction becomes a 90~mm wide collimated beam.

%%%%%%%%%%%%%%%%%%%%%%%%%%%%%%%%%%%%%%%%%%%%%%%%%%%%%%%%%%%%%%%%%%%%

\subsection{VPH grating cross-disperser}

After the cylindrical beam expander, a VPH grating cross-disperses the spectrum. The grating we selected is custom, with a line density of 1650 lines/mm. This grating is operated in Littrow configuration at our design wavelength $\lambda_\mathrm{des}=1083$~nm, in order to maximize diffraction efficiency. The angle of incidence is $63.3^\circ$. The clear aperture of the grating is rectanglar, with dimensions of around 200~mm by 40~mm. The diffraction efficiency is roughly 65\% across our spectrograph's wavelength range.

%%%%%%%%%%%%%%%%%%%%%%%%%%%%%%%%%%%%%%%%%%%%%%%%%%%%%%%%%%%%%%%%%%%%

\subsection{Camera and detector}

\begin{figure}[H]
\centering\includegraphics[width=0.75\textwidth]{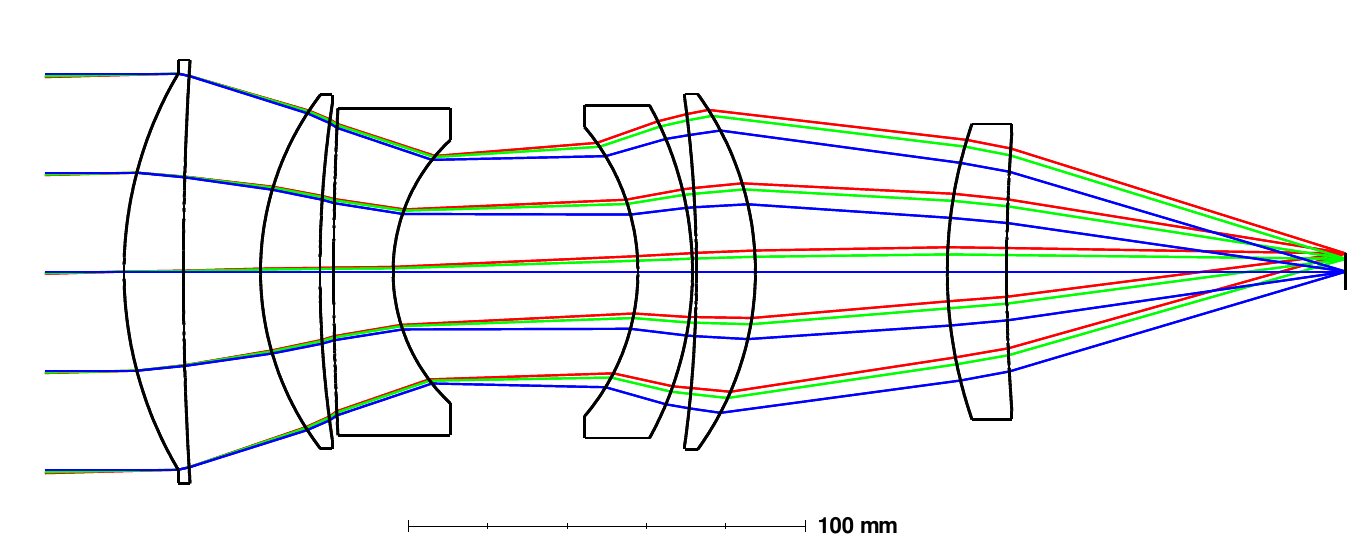}
\caption{Camera used in VIPER. The camera is a double Gauss design and is comprised of six spherical lenses.}\label{fig:Zemax_camera}
\end{figure}

\begin{figure}[H]
\centering\includegraphics[width=0.6\textwidth]{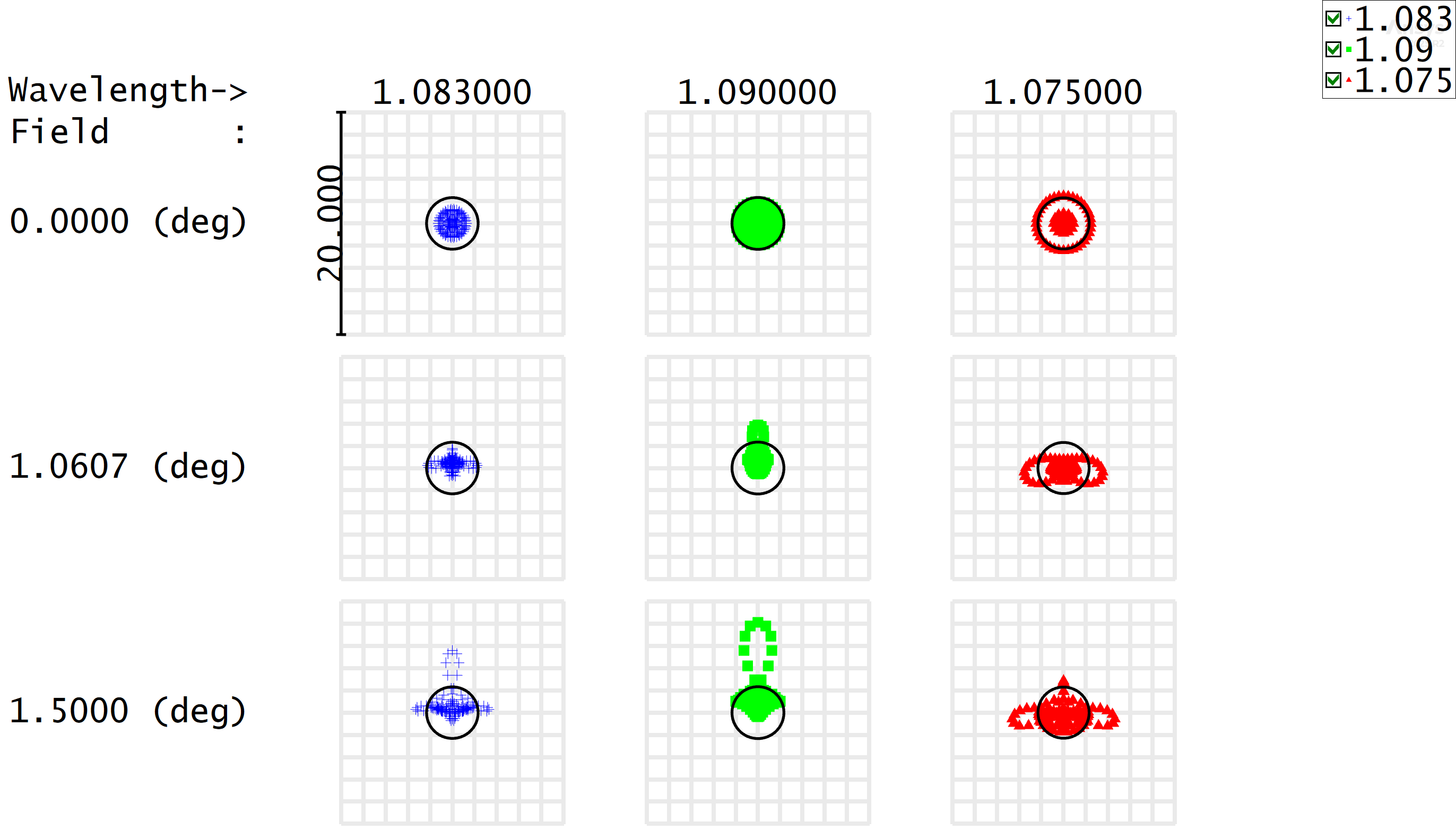}
\caption{Spot diagram for camera used in VIPER. The scale bar is 20 \textmu m. Wavelengths are in \textmu m. The black circles denote the Airy radius.}\label{fig:Cam_MatrixSpotDiagram}
\end{figure}

Figure~\ref{fig:Zemax_camera} shows the camera used in VIPER. The camera comprises six spherical lenses and is based on a double Gauss design. It has a focal length of $f_\mathrm{cam}=175.70$~mm and was designed with a focal ratio of f/1.757. The fast focal ratio led us to use a double Gauss design. The camera has near-diffraction-limited performance, as seen in Figure~\ref{fig:Cam_MatrixSpotDiagram}. Near-diffraction-limited imaging performance is necessary in our case. It is crucial that the camera optic in a VIPA spectrograph is nearly aberration-free, so that the spectrograph can operate at the designed resolving power\cite{Aryana2026}.

The focal length of the camera was selected based on the detector. We selected a sampling such that one wavelength resolution element (3.61~pm) spans three pixels, which is above Nyquist sampling. The detector we are using is based on the Sony IMX990 InGaAs CMOS, which has $1296\times1032$ pixels and a 5 \textmu m pixel pitch. The quantum efficiency of this detector is around 73\% across our spectrograph's wavelength range.

%%%%%%%%%%%%%%%%%%%%%%%%%%%%%%%%%%%%%%%%%%%%%%%%%%%%%%%%%%%%%%%%%%%%
%%%%%%%%%%%%%%%%%%%%%%%%%%%%%%%%%%%%%%%%%%%%%%%%%%%%%%%%%%%%%%%%%%%%

\subsection{Simulated performance}

\begin{figure}[H]
    \centering
    \begin{subfigure}{0.46\textwidth}
        \centering
        \includegraphics[width=\textwidth]{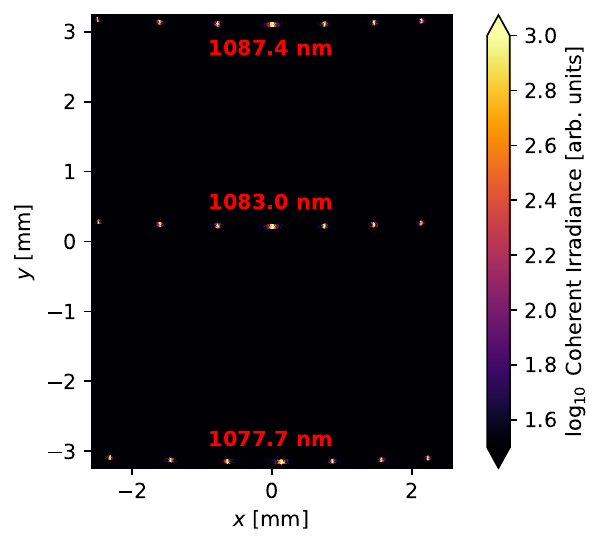}
        \caption{Zemax ray tracing simulation of three wavelengths.}
        \label{fig:NSC_wav_min_des_max}
    \end{subfigure}
    \begin{subfigure}{0.53\textwidth}
        \centering
        \includegraphics[width=\textwidth]{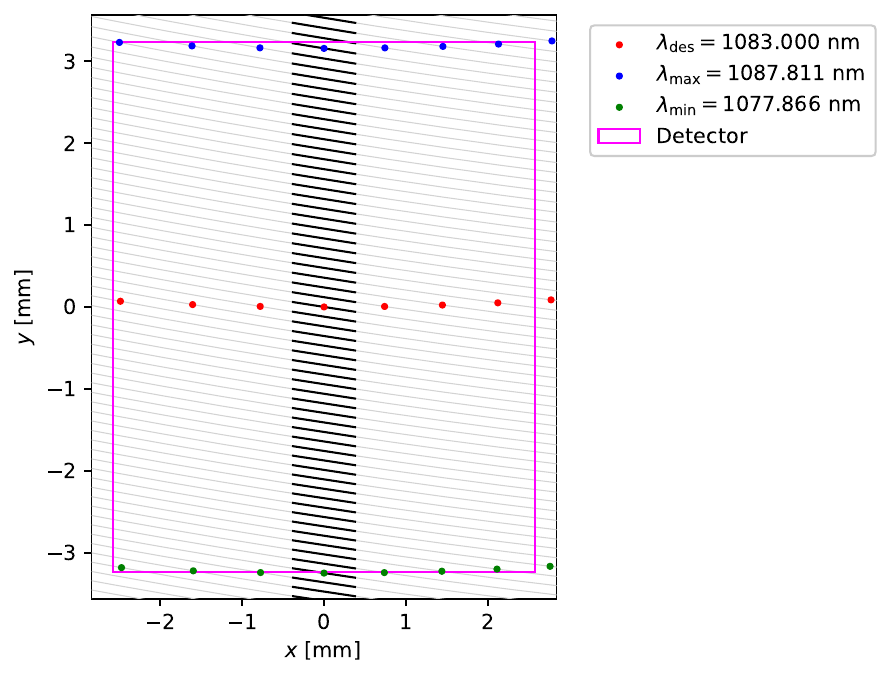}
        \caption{Detector footprint as computed from analytic relations.}
        \label{fig:footprint_detector_rotated}
    \end{subfigure}
    \caption{The left subfigure shows a view of the detector following a ray tracing simulation of VIPER's optical design in Zemax OpticStudio non-sequential mode, using three different wavelengths. The right subfigure shows the detector footprint, as computed from analytic relations\cite{Leung2025}. The orders are indicated by the gray curves. The black curves indicate unique wavelengths.}
    \label{fig:detector_wav_min_des_max}
\end{figure}

We validated our optical design using ray tracing simulations in Zemax OpticStudio non-sequential mode. Figure~\ref{fig:NSC_wav_min_des_max} shows a view of the detector following a ray tracing simulation with three wavelengths. Each wavelength shows up in around seven different orders. The light is distributed between these orders. Figure~\ref{fig:footprint_detector_rotated} shows the expected detector footprint, as computed from analytic relations\cite{Leung2025}. The VIPA dispersion direction is from right to left, while the cross-dispersion direction is from bottom to top. Figures~\ref{fig:NSC_wav_min_des_max} and \ref{fig:footprint_detector_rotated} appear similar and show that VIPER has a wavelength range of around 10~nm.

\begin{figure}[H]
\centering\includegraphics[width=0.5\textwidth]{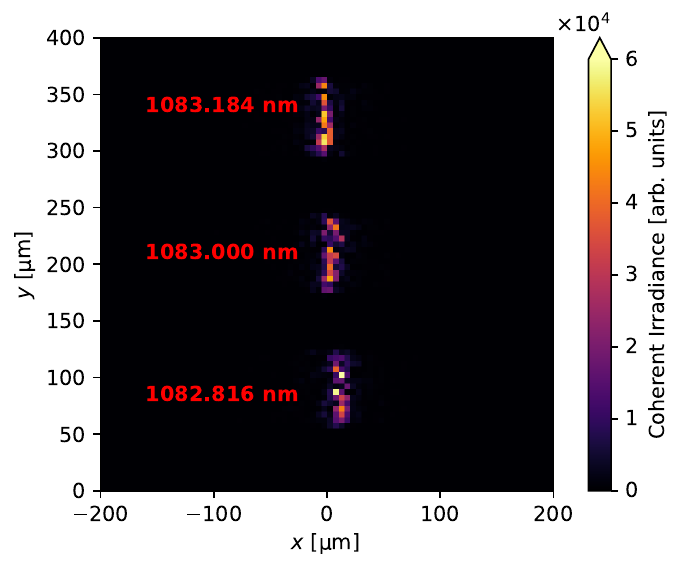}
\caption{Ray tracing simulation of VIPER's optical design, with three wavelengths in three adjacent orders, in Zemax OpticStudio non-sequential mode. Adjacent orders are well-separated.}\label{fig:NSC_wav_des_adjorders}
\end{figure}

Figure~\ref{fig:NSC_wav_des_adjorders} shows a ray tracing simulation of VIPER's optical design, with three wavelengths in three adjacent orders. From this ray trace, we can see that adjacent orders are well-separated and that there is adequate cross-dispersion.

\begin{figure}[H]
    \centering
    \begin{subfigure}{0.49\textwidth}
        \centering
        \includegraphics[width=\textwidth]{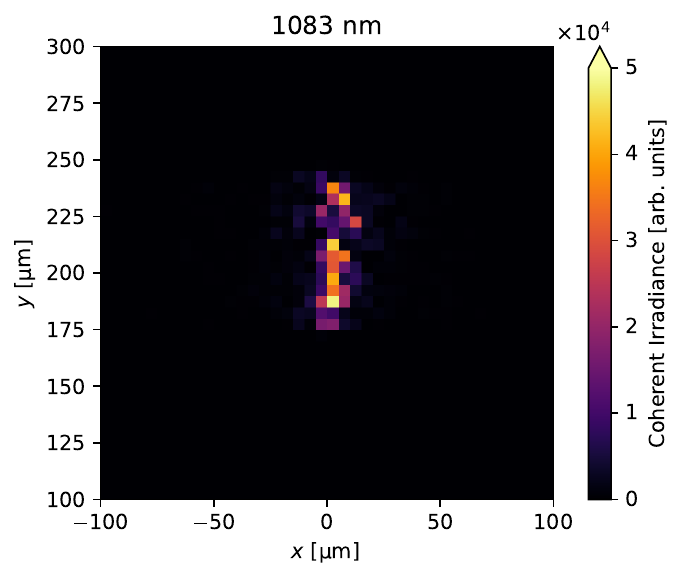}
        \caption{Point spread function at 1083~nm.}
        \label{fig:NSC_wav_des}
    \end{subfigure}
    \begin{subfigure}{0.49\textwidth}
        \centering
        \includegraphics[width=\textwidth]{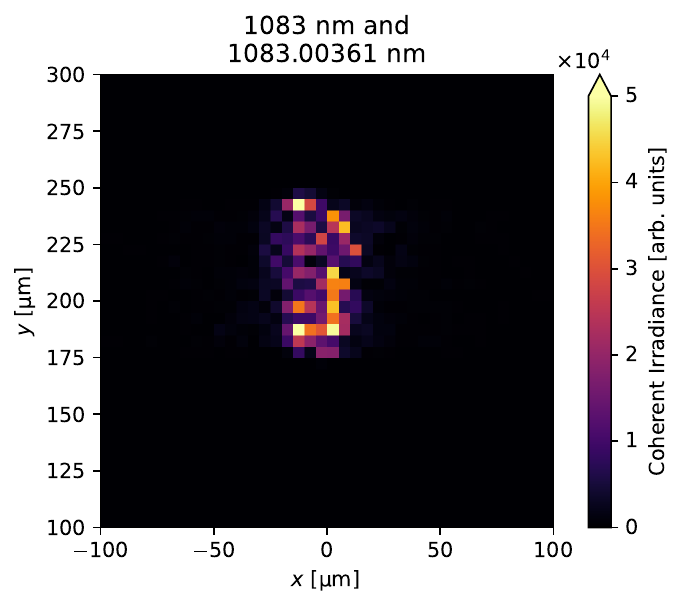}
        \caption{Point spread functions at 1083~nm and 1083.00361~nm.}
        \label{fig:NSC_wav_des_res}
    \end{subfigure}
    \caption{Point spread functions of VIPER at 1083~nm and 1083.00361~nm, from ray tracing simulations in Zemax OpticStudio non-sequential mode. These two wavelengths are separated by the wavelength resolution element, 3.61~pm, corresponding to a resolving power of 300,000. The point spread functions shown here are for the central diffraction order, closest to $x=0$.}
    \label{fig:detector_PSF}
\end{figure}

\begin{figure}[H]
\centering\includegraphics[width=0.55\textwidth]{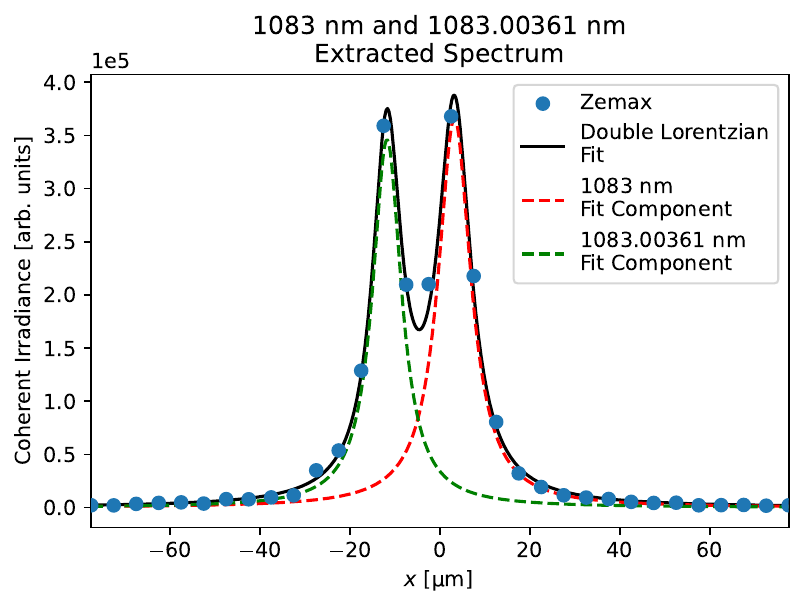}
\caption{Extracted spectrum of Figure~\ref{fig:NSC_wav_des_res}, shown by the blue points. The black curve is a fit to the blue points using a double Lorentzian model. Each Lorentzian is shown by the red and green dashed curves. The 3.61~pm wavelength resolution element is clearly resolved by VIPER.}\label{fig:NSC_wav_des_res_fit}
\end{figure}

Figure \ref{fig:detector_PSF} shows the point spread functions of VIPER at 1083~nm and 1083.00361~nm, from ray tracing simulations in Zemax OpticStudio non-sequential mode. These two wavelengths are separated by 3.61~pm, which is the wavelength resolution element corresponding to a resolving power of 300,000 at 1083 nm. From Figure~\ref{fig:NSC_wav_des}, we can see that one wavelength resolution element spans around three pixels in the VIPA dispersion direction, as designed. The point spread function spans around 13 pixels in the cross-dispersion direction. Figure~\ref{fig:NSC_wav_des_res_fit} shows an extracted spectrum of Figure~\ref{fig:NSC_wav_des_res}, along with a double Lorentzian fit to the extracted spectrum. We can see that the 3.61~pm resolution element is clearly resolved by VIPER.

Table~\ref{tab:E2E_efficiency} shows an estimate of the end-to-end instrument efficiency of VIPER. In this estimate, we have accounted for a front-end fiber injection efficiency of 84.4\%, assuming a seeing FWHM of $1.4''$, which is roughly the effective median seeing at the Tillinghast Telescope at our design wavelength $\lambda_\mathrm{des}=1083$ nm. We estimate the pupil slicer's total transmission to be above 90\%. This number was based on a pupil slicer made for the EXPRES\cite{Jurgenson2016} spectrograph at the Lowell Discovery Telescope, which our pupil slicer is based on (see Section \ref{sec:slicer}). The VIPA's total transmission is 78.1\%, as computed using the method described in Ref.~\citenum{Leung2026}. Also accounting for the VPH grating diffraction efficiency, coatings, and the detector quantum efficiency, we estimate that the end-to-end instrument efficiency of VIPER is around 27.6\%. This number is higher than most conventional high-resolution astronomical spectrographs, which have efficiencies of around 5--10\%. 

\begin{table}[H]
\centering
\begin{tabular}{ | m{0.45\textwidth} | m{0.15\textwidth}| } 
  \hline
  Component & Transmission \\ 
  \hline
  \hline
  Front-end fiber injection efficiency ($1.4''$ seeing) & $84.4\%$ \\
  Pupil slicer total transmission & $\gtrsim90\%$ \\
  VIPA total transmission & $78.1\%$ \\
  VPH grating diffraction efficiency & $65\%$ \\
  Coatings (30 surfaces) & $98.2\%$ \\
  Detector quantum efficiency & $73\%$ \\
  \hline
  Total & $27.6\%$ \\
  \hline
\end{tabular}
\caption{VIPER's estimated end-to-end efficiency at design wavelength $\lambda_\mathrm{des}=1083$ nm.}\label{tab:E2E_efficiency}
\end{table}
\section{Telescope, Front-End Module, and Pupil Slicer}\label{sec:telescope_FEM_PS}

\subsection{Telescope and front-end module}\label{sec:telescope_FEM}

\begin{figure}[H]
\centering\includegraphics[width=0.6\textwidth]{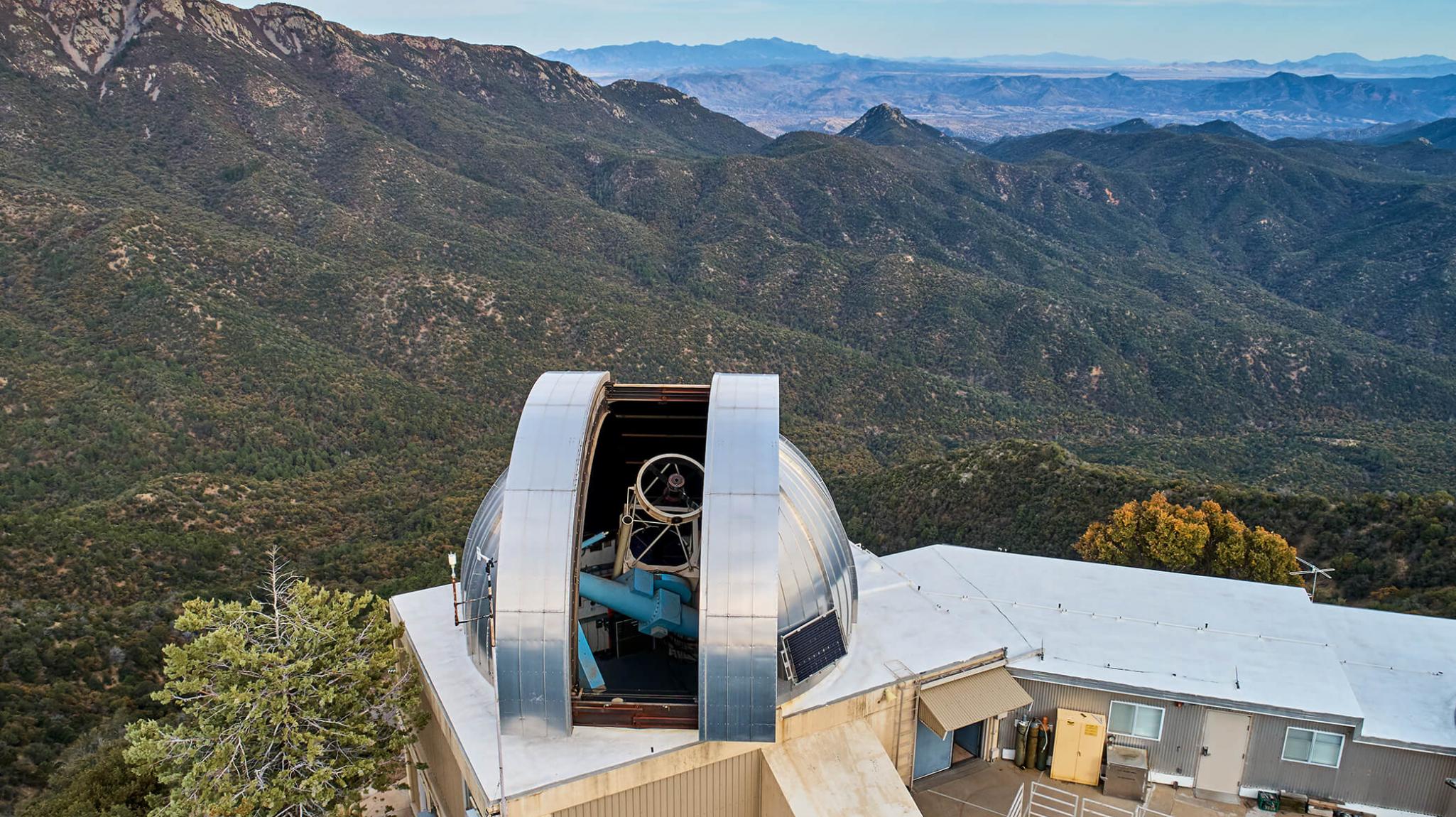}
\caption{Photo of the 1.5 m Tillinghast Telescope at the Fred Lawrence Whipple Observatory on Mount Hopkins, Arizona, USA.}\label{fig:Tillinghast}
\end{figure}

The Tillinghast Telescope is a Cassegrain reflector located at the FLWO Ridge on Mount Hopkins, Arizona, USA. The FLWO Ridge is located at 7800~ft above sea level and 800~ft below the peak of Mount Hopkins. The focal ratio of the Tillinghast Telescope is f/10. Its primary mirror has a diameter of 1.52~m, and its secondary mirror has a diameter of 0.40~m. Figure~\ref{fig:Tillinghast} shows a photo of the Tillinghast Telescope.

\begin{figure}[H]
\centering\includegraphics[width=0.6\textwidth]{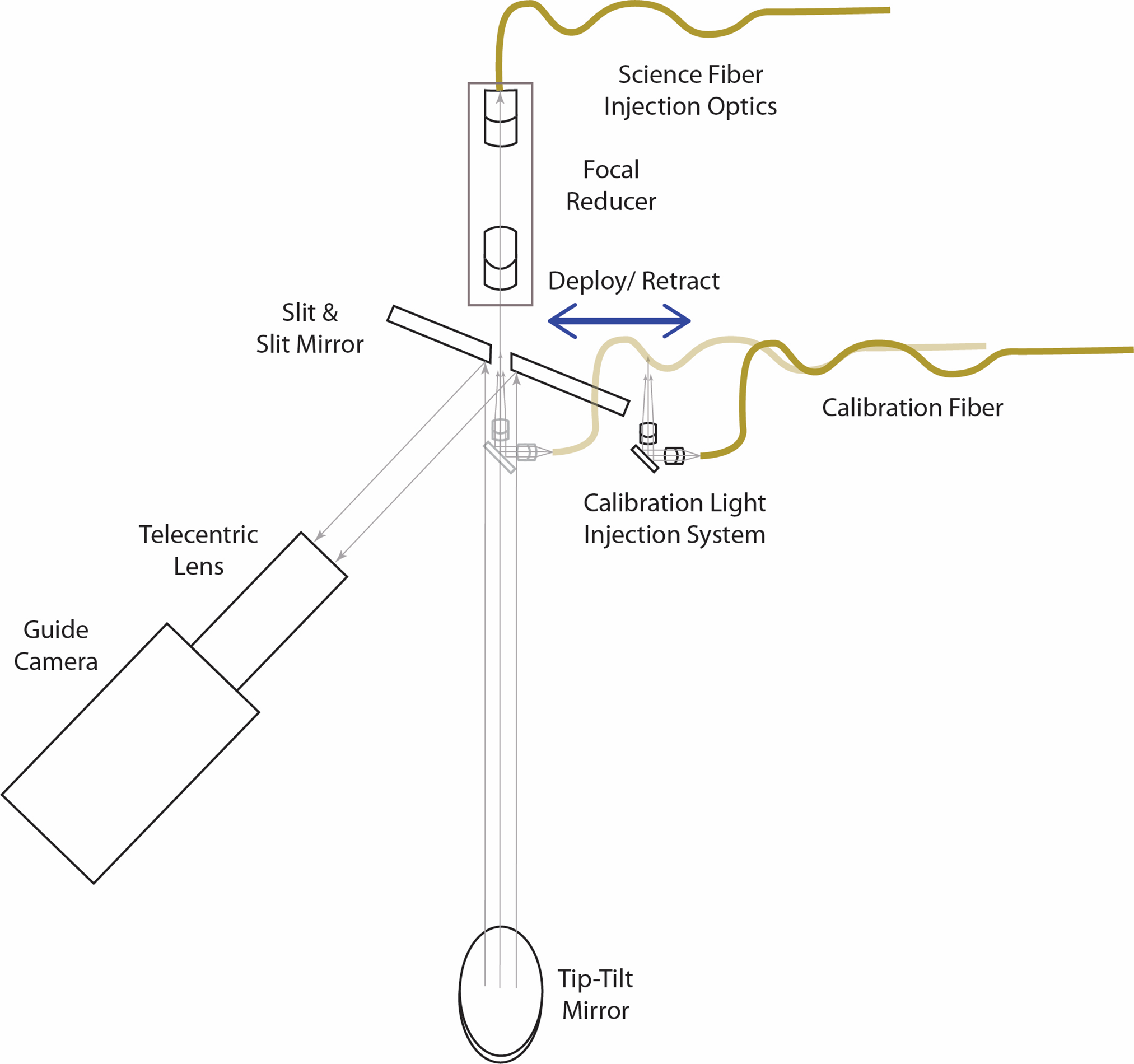}
\caption{Illustration of the front-end module for TRES. VIPER will utilize this same front-end.}\label{fig:TRES_NFE}
\end{figure}

VIPER will use the same front-end module as the one used for TRES. An illustration of this front-end module is shown in Figure~\ref{fig:TRES_NFE}. Light from the telescope is guided and reflected by a tip-tilt mirror into a focal reducer, which converts the telescope's f/10 beam to f/6 for injection into the science fiber. A mirror with a slit is placed before the focal reducer, reflecting light into an arm with a telecentric lens and a CMOS guiding camera. This camera looks at how light is being injected into the focal reducer, and the tip-tilt mirror is adjusted accordingly. The focal reducer consists of two bonded doublets. There is also a calibration light injection system that can be deployed or retracted, which feeds light from a calibration source into the science fiber. This calibration unit consists of two doublets and a mirror.

%%%%%%%%%%%%%%%%%%%%%%%%%%%%%%%%%%%%%%%%%%%%%%%%%%%%%%%%%%%%%%%%%%%%
%%%%%%%%%%%%%%%%%%%%%%%%%%%%%%%%%%%%%%%%%%%%%%%%%%%%%%%%%%%%%%%%%%%%

\subsection{Pupil slicer and double scrambler}\label{sec:slicer}

\begin{figure}[H]
\centering\includegraphics[width=\textwidth]{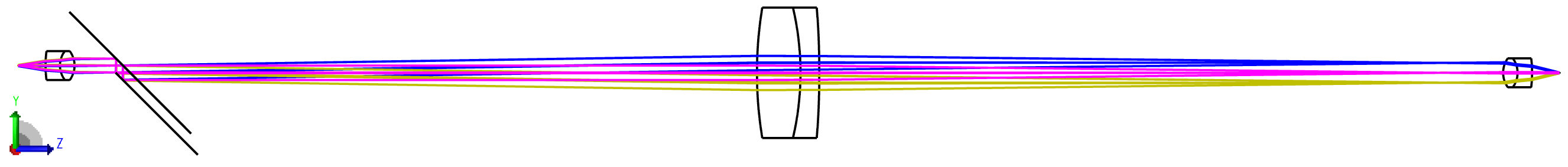}
\caption{Preliminary design of the combined mirror Bowen-Walraven pupil slicer and double scrambler for VIPER.}\label{fig:Slicer_3DLayout}
\end{figure}

\begin{figure}[H]
    \centering
    \begin{subfigure}{0.37\textwidth}
        \centering
        \includegraphics[width=\textwidth]{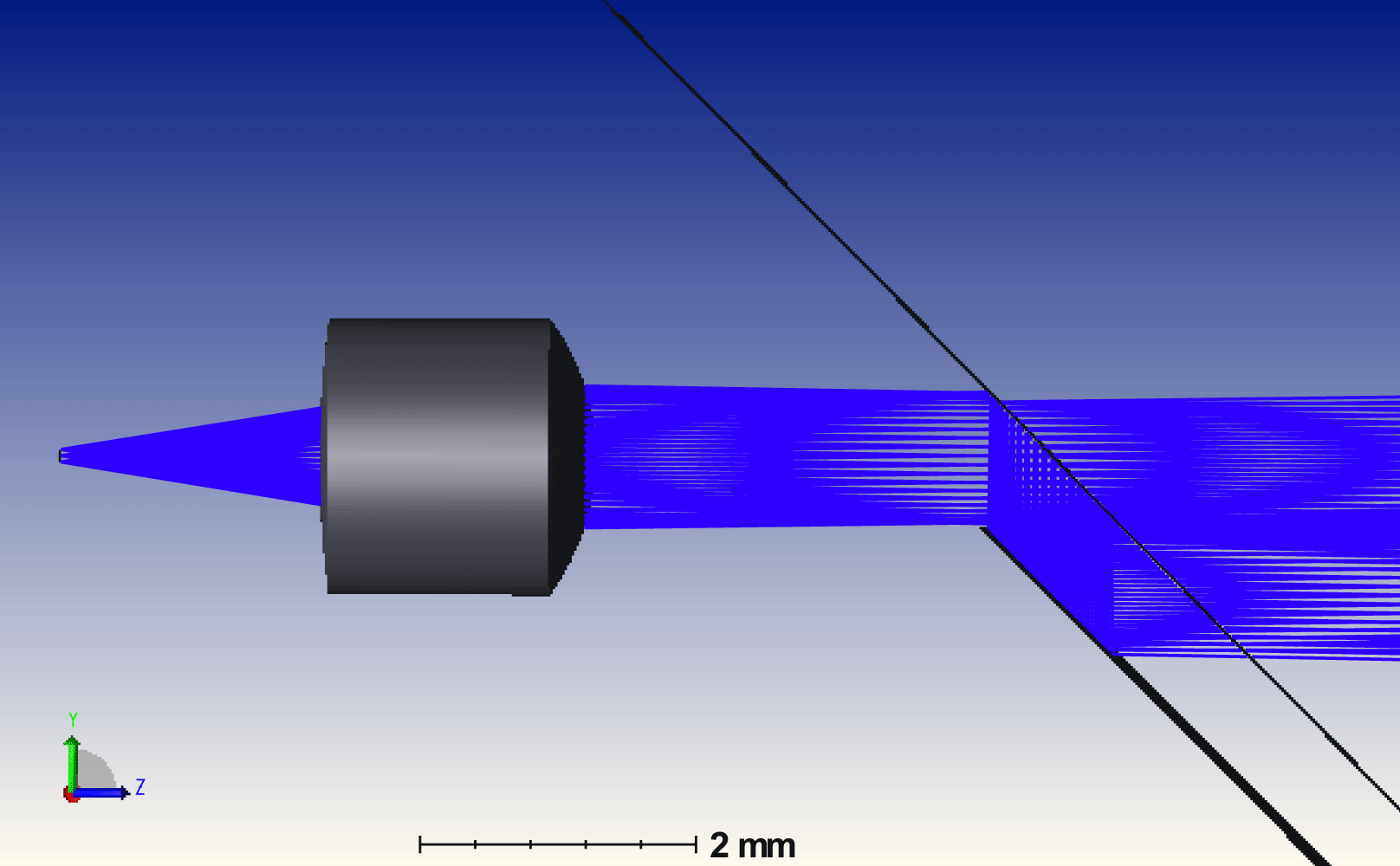}
        \caption{Side view.}
    \end{subfigure}
    \begin{subfigure}{0.27\textwidth}
        \centering
        \includegraphics[width=\textwidth]{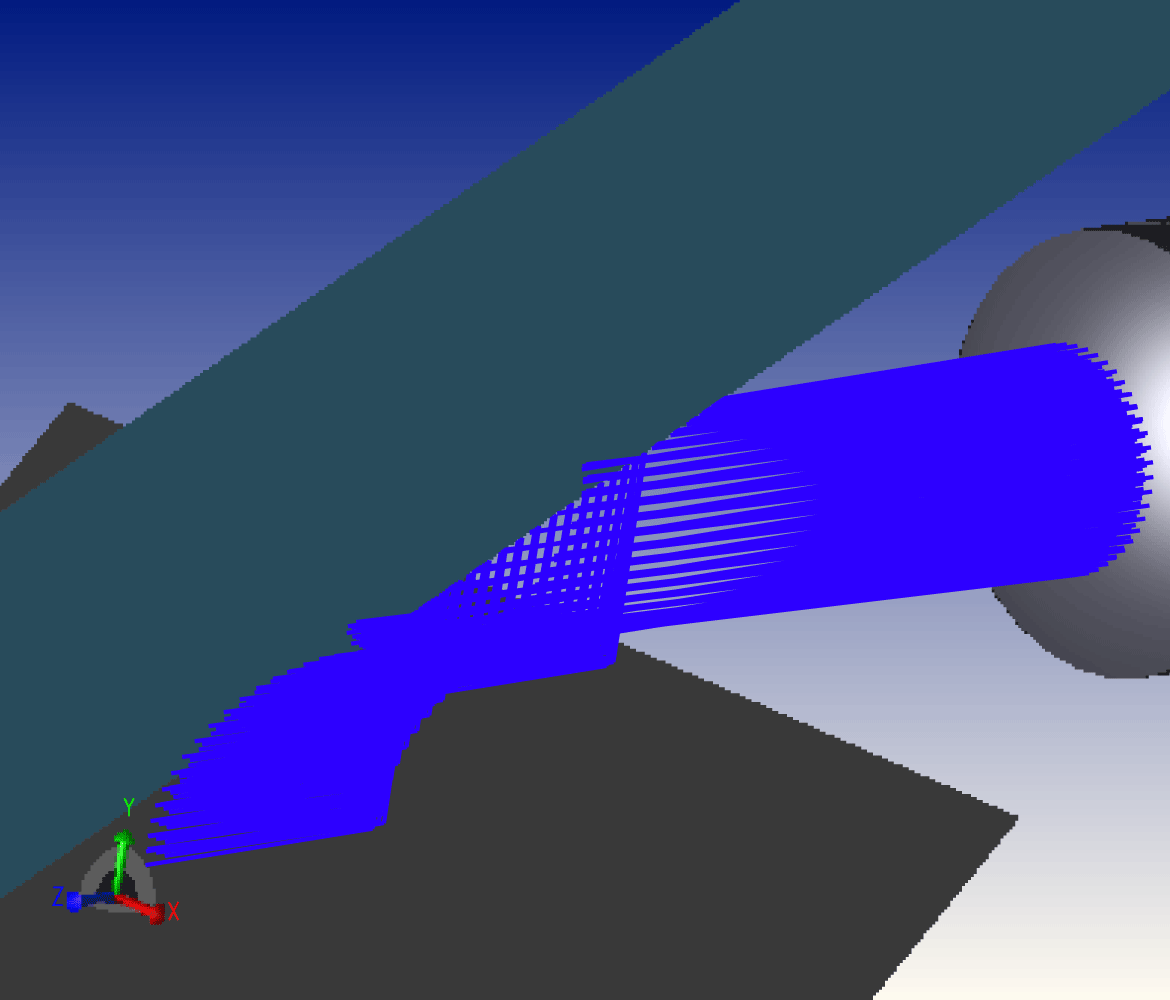}
        \caption{View with mirrors.}
    \end{subfigure}
    \begin{subfigure}{0.27\textwidth}
        \centering
        \includegraphics[width=\textwidth]{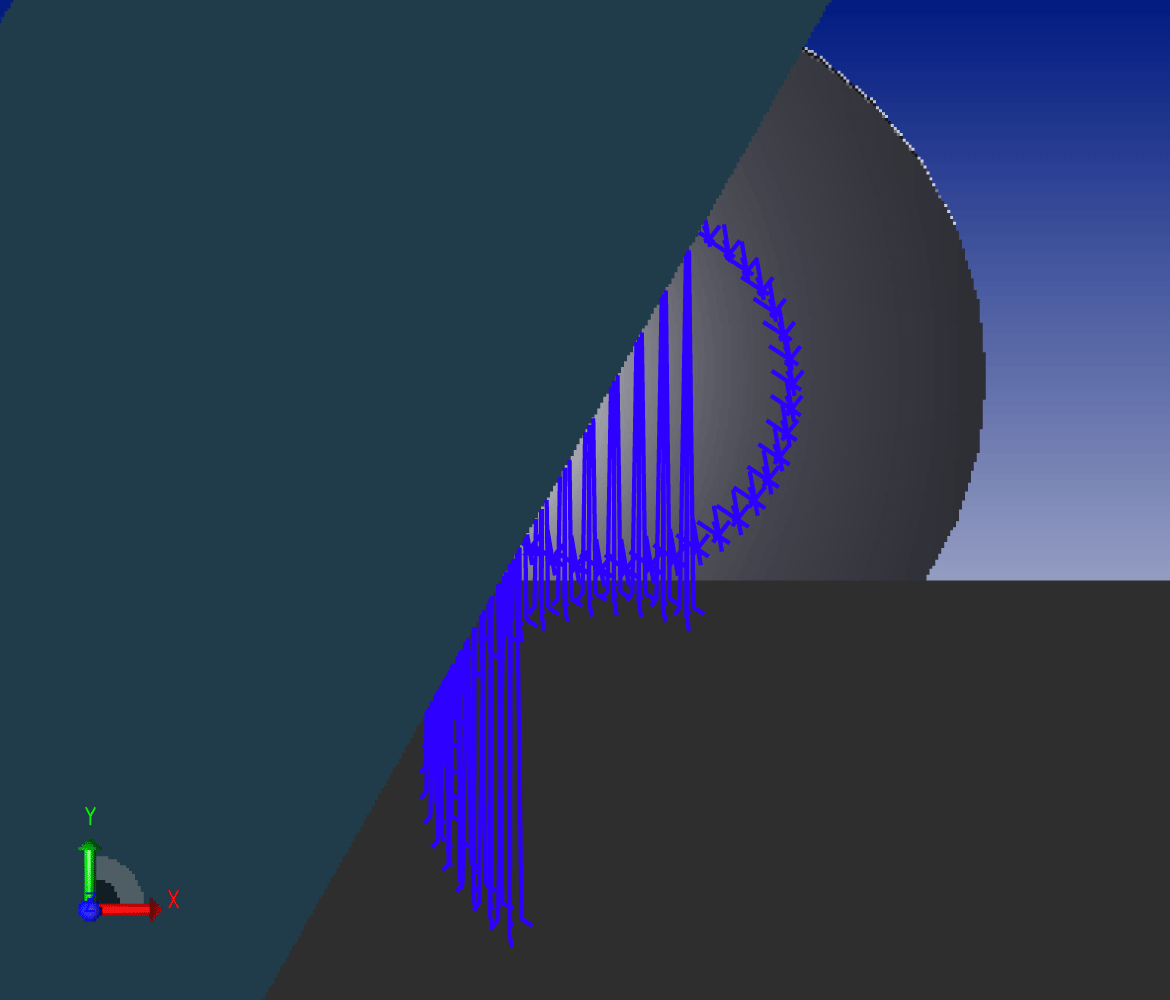}
        \caption{Front view.}
    \end{subfigure}
    \caption{Close-up views of mirror Bowen-Walraven pupil slicer seen in Figure~\ref{fig:Slicer_3DLayout}.}
    \label{fig:Slicer_ShadedModel}
\end{figure}

\begin{figure}[H]
    \centering
    \begin{subfigure}{0.66\textwidth}
        \centering
        \includegraphics[width=0.7\textwidth]{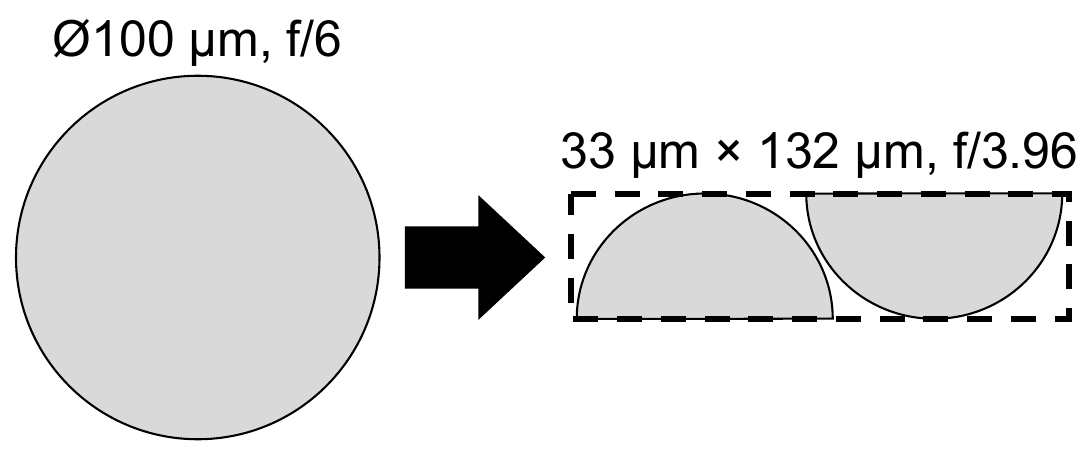}
        \caption{Illustration of fiber slicing scheme.}
    \end{subfigure}
    \begin{subfigure}{0.33\textwidth}
        \centering
        \includegraphics[width=0.6\textwidth]{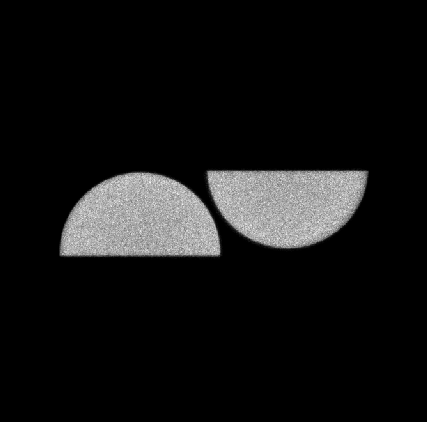}
        \caption{Zemax Geometric Image Analysis.}
    \end{subfigure}
    \caption{(Left) Illustration of fiber slicing scheme; (Right) corresponding Zemax OpticStudio Geometric Image Analysis simulation of the pupil slicer and double scrambler in Figure~\ref{fig:Slicer_3DLayout}.}
    \label{fig:slicer_explain}
\end{figure}

VIPER will utilize a combined pupil slicer and double scrambler. A mirror Bowen-Walraven pupil slicer design\cite{Tala2017,Avila2012} will be used. This combined slicer and double scrambler is based on the one used in EXPRES\cite{Jurgenson2016}. A preliminary design of this slicer and double scrambler is shown in Figures~\ref{fig:Slicer_3DLayout} and \ref{fig:Slicer_ShadedModel}. The slicer and double scrambler consists of three lens groups (each consisting of a doublet) and two mirrors. The two mirrors are placed at the pupil plane, one focal length after the first lens group. These mirrors slice the far-field of the Ø100~\textmu m circular fiber (f/6) from the TRES front-end into two semicircles and then recombine them to a roughly rectangular footprint. The next two lens groups then reimage the two semicircles onto a 33~\textmu m~$\times$~132~\textmu m rectangular fiber at f/3.96. The slicing scheme is illustrated in Figure~\ref{fig:slicer_explain}.

This slicer functions as a double scrambler, swapping the near-field and far-field of the Ø100~\textmu m circular fiber, and hence mitigating the second challenge discussed in Section~\ref{sec:MMF_VIPA_spec_challenges}. In addition to the pupil slicer and double scrambler, we will employ a mode scrambler (not to be confused with a double scrambler), also known as a ``fiber agitator'', to mitigate modal noise in the rectangular multimode fiber. We have developed a prototype mode scrambler\cite{Leung2025MS} for the GMT-Consortium Large Earth Finder (G-CLEF\cite{Szentgyorgyi2024}), a high-resolution fiber-fed spectrograph for the Magellan Clay Telescope and Giant Magellan Telescope (GMT), and we intend to use this kind of mode scrambler for VIPER as well. The development of the mode scrambler and combined pupil slicer and double scrambler of VIPER will happen in tandem with those of G-CLEF.

The main purpose of the pupil slicer is to transfer more etendue from the VIPA dispersion direction to the cross-dispersion direction. Recall that the rectangular fiber is oriented such that its shorter side is parallel to the VIPA dispersion direction and its longer side is parallel to the cross-dispersion direction. There is a limit to how much etendue can be coupled into a VIPA in a particular configuration\cite{Leung2026}, and so we put more etendue in the cross-dispersion direction instead. This increased etendue in the cross-dispersion direction is accounted for using the cylindrical beam expander, which effectively increases the focal length of the collimator along the cross-dispersion direction. Another way to think about this is that the cylindrical beam expander allows for more etendue to be accommodated in the cross-dispersion direction. This combination of a pupil slicer, a rectangular fiber, and a cylindrical beam expander together allows for a cross-dispersed multimode fiber-fed VIPA spectrograph design with high throughput.
\section{Conclusions and Next Steps}

We are developing VIPER: a high-resolution, multimode fiber-fed, narrowband, cross-dispersed, seeing-limited VIPA spectrograph for the 1.5~m Tillinghast Telescope at FLWO on Mount Hopkins, Arizona, USA. VIPER is matched to the Tillinghast Telescope's 100~\textmu m diameter circular multimode optical fiber feed at f/6. VIPER is specifically designed to probe exoplanet atmospheric escape through the metastable helium triplet line at 1083~nm.

In this work, we presented the optical design of VIPER, which is specifically optimized for high throughput, which, in a VIPA spectrograph, is more challenging with a multimode fiber feed than a single-mode fiber feed because of the larger etendue. The use of a multimode fiber feed gives rise to several challenges, including the need for more cross-dispersion, the impact of variations in fiber illumination, and the issue of VIPA coupling loss. Our solution to these challenges is to employ a pupil slicer, a rectangular fiber feed, and a cylindrical beam expander. Our resulting optical design is compact, fitting inside a square that roughly measures 50~cm by 50~cm. Our design has been validated using ray tracing simulations in Ansys Zemax OpticStudio non-sequential mode. The current design of VIPER has a resolving power of 300,000 over a 10~nm bandpass about the 1083~nm design wavelength. The estimated end-to-end efficiency of VIPER is around 27.6\%, which is higher than that of most conventional high-resolution grating-based astronomical spectrographs.

At the moment, we have started the process of procuring some components for VIPER. We have the VIPA and the detector in our lab, and we have conducted some experiments to investigate the temperature dependence of the VIPA. We are in talks with vendors about the other specialized components. We are planning to hold a design review for VIPER later this year. We aim for VIPER to be on-sky by 2027--2028.

%%%%%%%%%%%%%%%%%%%%%%%%%%%%%%%%%%%%%%%%%%%%%%%%%%%%%%%%%%%%%%%%%%%%
%%%%%%%%%%%%%%%%%%%%%%%%%%%%%%%%%%%%%%%%%%%%%%%%%%%%%%%%%%%%%%%%%%%%

\acknowledgments % equivalent to \section*{ACKNOWLEDGMENTS}       
 
We thank Surangkhana Rukdee, Daniel Sablowski, Kiumars Aryana, Joseph Zajac, Matthew Ashby, Daniel Fabricant, Dimitar Sasselov, and Liam Connor for helpful discussions. \mbox{M.C.H.~Leung} gratefully acknowledges the support of the Natural Sciences and Engineering Research Council of Canada (NSERC) through an NSERC Postgraduate Scholarship – Doctoral (PGS D).

% References
\bibliography{report} % bibliography data in report.bib

\begin{thebibliography}{10}

\bibitem{Schroeder}
Schroeder, D.~J.,  [{\em {Astronomical Optics}}{\nolinebreak\hspace{0.1em}]}, Academic Press, 2~ed. (2000).

\bibitem{Shirasaki1996}
Shirasaki, M., ``{Large angular dispersion by a virtually imaged phased array and its application to a wavelength demultiplexer},'' {\em Optics Letters}~{\bf 21}(5),  366 (1996).

\bibitem{Leung2025}
Leung, M. C.~H., Charbonneau, D., Szentgyorgyi, A., Jurgenson, C., MacLeod, M., Rukdee, S., Vissapragada, S., Nail, F., Zajac, J., and Dupree, A.~K., ``{VIPER: a high-resolution multimode fiber-fed VIPA spectrograph concept for characterizing exoplanet atmospheric escape},'' in [{\em Techniques and Instrumentation for Detection of Exoplanets XII}{\nolinebreak\hspace{0.1em}]},  {\em Proc. SPIE} {\bf 13627},  136271P (2025).

\bibitem{Zhu2023}
Zhu, X., Lin, D., Zhang, Z., Xie, X., and He, J., ``{Dispersion Characteristics of the Multi-mode Fiber-fed VIPA Spectrograph},'' {\em The Astronomical Journal}~{\bf 165}(6),  228 (2023).

\bibitem{SalehTeich}
Saleh, B. E.~A. and Teich, M.~C.,  [{\em {Fundamentals of Photonics}}{\nolinebreak\hspace{0.1em}]}, John Wiley \& Sons, Inc. (1991).

\bibitem{Bourdarot2018}
Bourdarot, G., Le~Coarer, E., Mouillet, D., Jocou, L., Rabou, P., Correia, J.-J., Bonfils, X., Stadler, E., Carlotti, A., Forveille, T., Vigan, A., Artigau, E., Doyon, R., Vallée, P., and Magnard, Y., ``{Experimental test of a 40 cm-long R=100 000 spectrometer for exoplanet characterisation},'' in [{\em Ground-based and Airborne Instrumentation for Astronomy VII}{\nolinebreak\hspace{0.1em}]},  {\em Proc. SPIE} {\bf 10702},  107025Y (2018).

\bibitem{Carlotti2022}
Carlotti, A., Bidot, A., Mouillet, D., Correia, J.-J., Jocou, L., Curaba, S., Delboulbé, A., Le~Coarer, E., Rabou, P., Bourdarot, G., Forveille, T., Bonfils, X., Vasisht, G., Mawet, D., Burruss, R.~S., Oppenheimer, R., Doyon, R., Artigau, E., and Vallée, P., ``{On-sky demonstration at Palomar Observatory of the near-IR, high-resolution VIPA spectrometer},'' in [{\em Ground-based and Airborne Instrumentation for Astronomy IX}{\nolinebreak\hspace{0.1em}]},  {\em Proc. SPIE} {\bf 12184},  52 (2022).

\bibitem{Leung2026}
Leung, M. C.~H., Szentgyorgyi, A., Jurgenson, C., and Charbonneau, D., ``{Coupling loss and transmission in a multimode fiber-fed Virtually Imaged Phased Array (VIPA)}.'' {\hypersetup{hidelinks}\url{https://doi.org/10.48550/arXiv.2607.02806}}, DOI: 10.48550/arXiv.2607.02806 (2026).

\bibitem{Oklopi2018}
Oklopčić, A. and Hirata, C.~M., ``{A New Window into Escaping Exoplanet Atmospheres: 10830 {\AA} Line of Helium},'' {\em The Astrophysical Journal}~{\bf 855}(1),  L11 (2018).

\bibitem{Strader2015}
Strader, J., Dupree, A.~K., and Smith, G.~H., ``{THE 10830 {\AA} HELIUM LINE AMONG EVOLVED STARS IN THE GLOBULAR CLUSTER M4},'' {\em The Astrophysical Journal}~{\bf 808}(2),  124 (2015).

\bibitem{Dupree1986}
Dupree, A.~K., ``{Mass Loss from Cool Stars},'' {\em Annual Review of Astronomy and Astrophysics}~{\bf 24}(1),  377–420 (1986).

\bibitem{Cooke2022}
Cooke, R.~J., Noterdaeme, P., Johnson, J.~W., Pettini, M., Welsh, L., Peroux, C., Murphy, M.~T., and Weinberg, D.~H., ``{Primordial Helium-3 Redux: The Helium Isotope Ratio of the Orion Nebula*},'' {\em The Astrophysical Journal}~{\bf 932}(1),  60 (2022).

\bibitem{Nail2025}
Nail, F., MacLeod, M., Oklopčić, A., Gully-Santiago, M., Morley, C.~V., and Zhang, Z., ``Cold dayside winds shape large leading streams in evaporating exoplanet atmospheres,'' {\em Astronomy \& Astrophysics}~{\bf 695},  A186 (2025).

\bibitem{Owen2022}
Owen, J.~E., Murray-Clay, R.~A., Schreyer, E., Schlichting, H.~E., Ardila, D., Gupta, A., Loyd, R. O.~P., Shkolnik, E.~L., Sing, D.~K., and Swain, M.~R., ``{The fundamentals of Lyman $\alpha$ exoplanet transits},'' {\em Monthly Notices of the Royal Astronomical Society}~{\bf 518}(3),  4357–4371 (2022).

\bibitem{Fulton2017}
Fulton, B.~J., Petigura, E.~A., Howard, A.~W., Isaacson, H., Marcy, G.~W., Cargile, P.~A., Hebb, L., Weiss, L.~M., Johnson, J.~A., Morton, T.~D., Sinukoff, E., Crossfield, I. J.~M., and Hirsch, L.~A., ``{The California-Kepler Survey. III. A Gap in the Radius Distribution of Small Planets*},'' {\em The Astronomical Journal}~{\bf 154}(3),  109 (2017).

\bibitem{Mazeh2016}
Mazeh, T., Holczer, T., and Faigler, S., ``{Dearth of short-period Neptunian exoplanets: A desert in period-mass and period-radius planes},'' {\em Astronomy \& Astrophysics}~{\bf 589},  A75 (2016).

\bibitem{Szentgyorgyi2007}
Szentgyorgyi, A.~H. and Furész, G., ``{Precision Radial Velocities for the Kepler Era},'' in [{\em The 3rd Mexico-Korea Conference on Astrophysics: Telescopes of the Future and San Pedro Mártir}{\nolinebreak\hspace{0.1em}]},  {\em Revista Mexicana de Astronomía y Astrofísica} {\bf 28},  129--133 (2007).

\bibitem{Leung2025MS}
Leung, M. C.~H., Jurgenson, C., Szentgyorgyi, A., Podgorski, W., Mueller, M., Rimalt, Y.~S., Zajac, J., Onyuksel, C., Durusky, D., and Doherty, P., ``{Crank-rocker optical fiber mode scrambler prototype for the GMT-Consortium Large Earth Finder (G-CLEF)},'' in [{\em Techniques and Instrumentation for Detection of Exoplanets XII}{\nolinebreak\hspace{0.1em}]},  {\em Proc. SPIE} {\bf 13627},  136271J (2025).

\bibitem{Epworth1979}
Epworth, R.~E., ``{Phenomenon of modal noise in fiber systems},'' in [{\em Optical Fiber Communication}{\nolinebreak\hspace{0.1em}]},   ThD1, OSA (1979).

\bibitem{Aryana2026}
Aryana, K., Bailey, D.~M., Woods, S.~I., and Fleisher, A.~J., ``{Bridging Theory and Experiment in Virtually Imaged Phased Array (VIPA) Spectrometers}.'' {\hypersetup{hidelinks}\url{https://arxiv.org/abs/2601.08589}}, DOI: 10.48550/ARXIV.2601.08589 (2026).

\bibitem{Jurgenson2016}
Jurgenson, C., Fischer, D., McCracken, T., Sawyer, D., Szymkowiak, A., Davis, A., Muller, G., and Santoro, F., ``{EXPRES: a next generation RV spectrograph in the search for earth-like worlds},'' in [{\em Ground-based and Airborne Instrumentation for Astronomy VI}{\nolinebreak\hspace{0.1em}]},  {\em Proc. SPIE} {\bf 9908},  99086T (2016).

\bibitem{Tala2017}
Tala, M., Vanzi, L., Avila, G., Guirao, C., Pecchioli, E., Zapata, A., and Pieralli, F., ``Two simple image slicers for high resolution spectroscopy,'' {\em Experimental Astronomy}~{\bf 43}(2),  167–176 (2017).

\bibitem{Avila2012}
Avila, G., Guirao, C., and Baader, T., ``High efficiency inexpensive 2-slices image slicers,'' in [{\em Ground-based and Airborne Instrumentation for Astronomy IV}{\nolinebreak\hspace{0.1em}]},  {\em Proc. SPIE} {\bf 8446},  84469M, SPIE (2012).

\bibitem{Szentgyorgyi2024}
Szentgyorgyi, A., Ben-Ami, S., Oh, J.~S., et~al., ``{Innovations in the design and construction of the GMT-Consortium Large Earth Finder (G-CLEF), a first-light instrument for the Giant Magellan Telescope (GMT)},'' in [{\em Ground-based and Airborne Instrumentation for Astronomy X}{\nolinebreak\hspace{0.1em}]},  {\em Proc. SPIE} {\bf 13096},  130960Z (2024).

\end{thebibliography}
\bibliographystyle{spiebib} % makes bibtex use spiebib.bst

\end{document}